\documentclass[11pt, a4paper]{article}

\usepackage{jcapmod}

\usepackage[latin1]{inputenc}
\usepackage[british]{babel}
\makeatletter\AtBeginDocument{\let\@elt\relax}\makeatother
\usepackage{amsmath}
\usepackage{amssymb}
\usepackage{physics}
\usepackage{amsthm}
\usepackage[]{graphicx}
\usepackage[]{subfigure}
\usepackage{tensor}
\usepackage{color}
\usepackage{cancel}
\usepackage{setspace}
\usepackage{fancyhdr}
\usepackage{hyperref}
\usepackage{amsfonts}
\usepackage{mathtools}
\usepackage{bm}
\usepackage{textalpha}

\numberwithin{equation}{section}

\usepackage[a4paper,margin=2.5cm]{geometry}

\renewcommand{\baselinestretch}{1.3}

\begin{document}
	
	\pagenumbering{roman}
	
	\begin{titlepage}
		
		\baselineskip=15.5pt \thispagestyle{empty}
		
		%%% TITLE %%%
		\begin{center}
			{\fontsize{20.74}{24}\selectfont \bfseries Uses of Complex Metrics in Cosmology}
		\end{center}
		
		\vspace{0.1cm}
		
		%%% AUTHORS %%%
		\begin{center}
			{\fontsize{12}{18}\selectfont Caroline Jonas,$^{1}$ Jean-Luc Lehners,$^{1}$ and Jerome Quintin$^{1,2,3}$}
		\end{center}
		
		%%% AFFILIATIONS %%%
		\begin{center}
			\vskip8pt
			\textsl{$^1$ Max Planck Institute for Gravitational Physics (Albert Einstein Institute),\\
				Am M\"uhlenberg 1, D-14476 Potsdam, Germany}
			\vskip6pt
			\textsl{$^2$ The Fields Institute for Research in Mathematical Sciences, University of Toronto,\\
				222 College St., Toronto, Ontario M5T 3J1, Canada}
			\vskip6pt
			\textsl{$^3$ Dept.~of Applied Mathematics and Waterloo Centre for Astrophysics, University of Waterloo,\\
				200 University Ave.~W., Waterloo, Ontario N2L 3G1, Canada}
		\end{center}
		
		\vspace{1.2cm}
		
		%%% ABSTRACT %%%
		\hrule
		\vspace{0.3cm}
		\noindent {\bf Abstract}\\[0.1cm]
		Complex metrics are a double-edged sword: they allow one to replace singular spacetimes, such as those containing a big bang, with regular metrics, yet they can also describe unphysical solutions in which quantum transitions may be more probable than ordinary classical evolution. In the cosmological context, we investigate a criterion proposed by Witten (based on works of Kontsevich \& Segal and of Louko \& Sorkin) to decide whether a complex metric is allowable or not. Because of the freedom to deform complex metrics using Cauchy's theorem, deciding whether a metric is allowable in general requires solving a complicated optimisation problem. We describe a method that allows one to quickly determine the allowability of minisuperspace metrics. This enables us to study the off-shell structure of minisuperspace path integrals, which we investigate for various boundary conditions. Classical transitions always reside on the boundary of the domain of allowable metrics, and care must be taken in defining appropriate integration contours for the corresponding gravitational path integral. Perhaps more surprisingly, we find that proposed quantum (`tunnelling') transitions from a contracting to an expanding universe violate the allowability criterion and may thus be unphysical. No-boundary solutions, by contrast, are found to be allowable, and moreover we demonstrate that with an initial momentum condition an integration contour over allowable metrics may be explicitly described in arbitrary spacetime dimensions. 
		
		\vskip10pt
		\hrule
		\vskip10pt
		
	\end{titlepage}
	
	\thispagestyle{empty}
	
	%%% TABLE OF CONTENTS %%%
	\setcounter{page}{2}
	\tableofcontents
	
	%\newpage
	\pagenumbering{arabic}
	%\setcounter{page}{1}
	%\clearpage

	%%%%%%%%%%%%%%%%%%%%%%%%%%%%%%%%%%%%%%%%%%%%%%%%%%%%%%%%%%%%%
	%%%%%%%%%%%%%%%%%%%%%%%%%%%%%%%%%%%%%%%%%%%%%%%%%%%%%%%%%%%%%
	
	%%% INTRODUCTION %%%
	
	\section{Introduction}
	\label{sec:introduction}
	
	Einstein's equations work just as well with complex-valued metrics as with real-valued ones. In loose analogy with quantum mechanics, where we work with a complex wave function, yet real observables, there seems to be no obstacle to use complex-valued metrics as long as the observable quantities remain real. In fact, analytically continued metrics have proven very useful, with the best-known examples linked to black hole physics: complexified black hole metrics provide the shortest route to calculating thermodynamic properties of black holes \cite{Gibbons:1976ue}.
	
	In cosmology, the main hope is that complex metrics may be able to regularise the big bang by replacing it with a non-singular geometry. In this vein, complex metrics may provide glimpses of quantum gravity, allowing us to see beyond the singularity that is all but unavoidable when working with Lorentzian metrics and stable matter fields \cite{Penrose:1964wq,Hawking:1966jv,Hawking:1967ju,Hawking:1970zqf,Hawking:1973uf}. Moreover, in the path integral approach to quantum gravity, where one sums over geometries, it is very natural to include complex metrics in the sum, for instance in order to find good integration contours \cite{Halliwell:1989dy}.
	
	Despite these appealing features, it seems clear that not all complex metrics can make physical sense. For example, Witten has discussed complex wormhole solutions with vanishing action \cite{Witten:2021nzp} -- it is clear that such geometries should not be included in a path integral, as they would have the same weighting as classical solutions. Hence there must be some criterion (or perhaps a whole set of criteria) that specifies which metrics are allowed and which not.
	
	This question has been discussed over many years (e.g., \cite{Ivashchuk:1987mnk,Louko:1995jw,Sorkin:2009ka,Visser:2017atf}) and sparked renewed interest recently. In a seminal work, Louko \& Sorkin discussed topology change in two dimensions and proposed that one should only allow complex geometries on which a real scalar field theory can be consistently defined \cite{Louko:1995jw}. By consistently defined, they meant that the path integral for a scalar field should be manifestly convergent on allowed geometries. Quite independently, Kontsevich \& Segal have considered the question of which background geometries are suitable for defining quantum field theories on them \cite{Kontsevich:2021dmb}. Their criterion may be seen as a generalisation of that of \cite{Louko:1995jw} to arbitrary dimensions and arbitrary quantum fields with local energy momentum tensors. Witten then proposed to investigate this same criterion in quantum gravity \cite{Witten:2021nzp}, noting that complex metrics that have proven useful, such as complexified black holes, satisfy said convergence conditions, while a number of pathological metrics, such as the complex wormholes mentioned before, do not.
	
	In the same spirit, we wish to continue an exploration of the Louko-Sorkin-Kontsevich-Segal-Witten (LSKSW) criterion in the context of early universe cosmology, in particular in minisuperspace models. In a preliminary study, one of us \cite{Lehners:2021mah} had noted that the LSKSW criterion may have significant consequences for gravitational integration contours and that no-boundary saddle points lie on the boundary of the allowed domain of metrics when the no-boundary wave function is defined in momentum space. Here we generalise the latter finding to arbitrary spacetime dimensions and also identify suitable integration contours. The boundary of the space of allowable metrics again plays a big role, just as it did in the work of Kontsevich \& Segal. Indeed, Lorentzian metrics all reside on the boundary, but we find that in minisuperspace path integrals, it can be non-trivial to reach this boundary along an integration contour. Crucial in such an analysis is the ability to identify which metrics are allowable and which not. This is a rather non-trivial task, and we discuss several methods that tackle this question. We also discuss examples of quantum transitions across (or rather `around') the big bang, from a contracting to an expanding universe. The examples we consider are found to be non-allowable, and they cast doubt on whether quantum bounces may be consistent. This is a striking example showing the impact that a restriction on complex metrics can have. 
	
	This work may also be seen as a refinement of our earlier work on cosmological amplitudes \cite{Jonas:2021xkx}, the take-home message being that semi-classical consistency is a powerful condition, and models that satisfy it should be given our serious attention. In the absence of laboratory experiments creating universes, mathematical consistency may yet lead us to the far reaches and the early times of our universe.

	The table of contents above provides the outline for our paper. A word on notation: we use Greek letters to denote spacetime indices regardless of the metric signature and for any spacetime dimensions. We also use units of $8\pi G_\mathrm{N}=1$ throughout, and in numerical computations we set the cosmological constant to $\Lambda=3$ unless stated otherwise.

	%%%%%%%%%%%%%%%%%%%%%%%%%%%%%%%%%%%%%%%%%%%%%%%%%%%%%%%%%%%%%%%%%%%%%%%%%%%%%%%%%%%%%%%%%%%%%%%%%%%%%%%
	
	\section{Metric allowability}
	\label{sec:allowableMetrics}
	
	\subsection{Pathological complex metrics}
	\label{subsec:noboundarynspheres}
	
	As recently emphasized by Witten \cite{Witten:2021nzp}, when considering the gravitational path integral it proves necessary to restrict the sum over complex metrics; many pathological examples can be constructed when one allows the metric to take any complex values. Here we will illustrate this point with an example familiar from the no-boundary proposal.
	
\begin{figure}[t]
		\centering
		\includegraphics[width=0.25\textwidth]{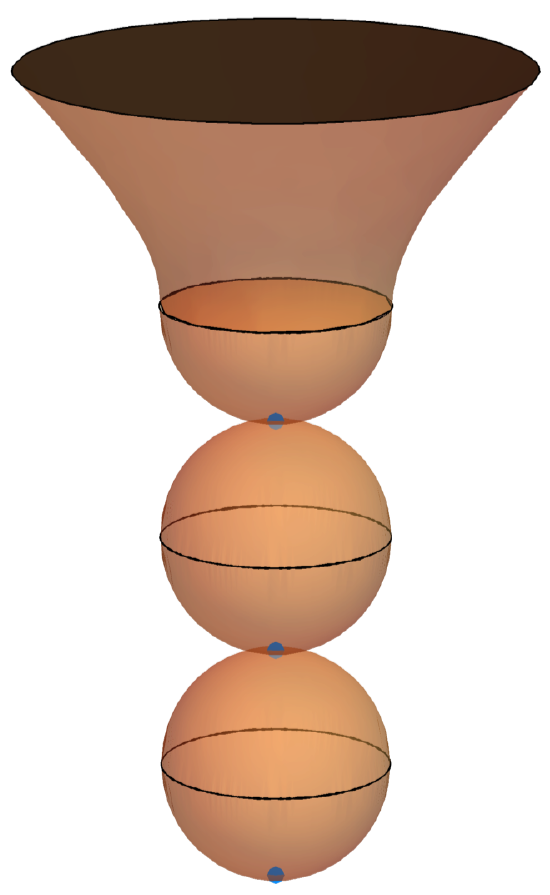}
		\hspace*{1cm}
		\includegraphics[width=0.55\textwidth]{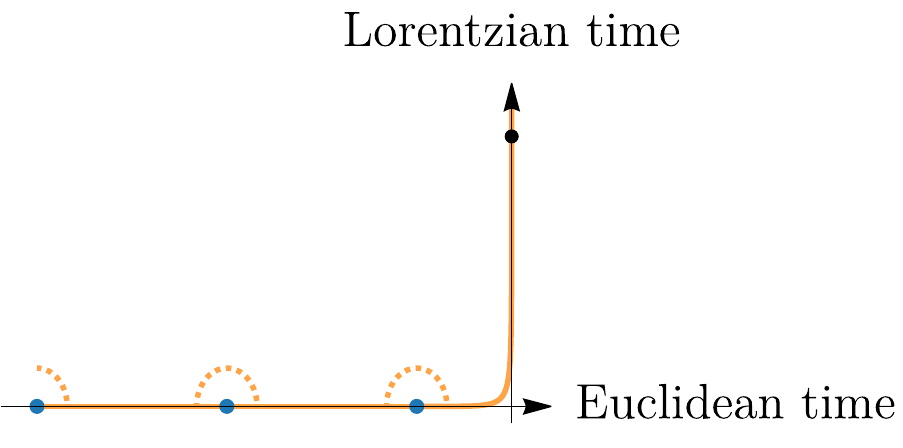}
		\caption{\textit{Left:} Spacetime depiction of the no-boundary solution with two additional spheres glued onto the original South Pole. `Time' evolves forwards from bottom to top: it is Euclidean when the geometries are spheres and Lorentzian when it becomes the hyperboloid. \textit{Right:} Graphical representation of the time path (orange curve) in the complex plane for $\tau$, where the horizontal axis denotes Euclidean time (real $\tau$, imaginary $t$) and where the vertical axis denotes Lorentzian time (imaginary $\tau$, real $t$). The curve changes direction at the moment of the Wick rotation (of $90^{\circ}$ in the complex plane). The blue dots indicate the south poles of the Euclidean spheres, and the black dot indicates the late-time Lorentzian de Sitter geometry.
			The dotted orange curves show possible deformations of the time path such that the zeroes of the scale factor are avoided by going into the Lorentzian direction.}
		\label{fig:no-boundaryManySpheres}
	\end{figure}

	The no-boundary proposal is a theory for the initial conditions of the universe, originally proposed by Hartle \& Hawking \cite{Hartle:1983ai}. It is formulated as a Euclidean path integral over compact and regular four-dimensional manifolds, whose only boundary is the present three-dimensional hypersurface. In the semi-classical limit, this path integral is approximated by the sum over its saddle point geometries (also referred to as \textit{instanton} geometries). These geometries are generally complex, making the no-boundary proposal an ideal testbed for a criterion of allowable complex metrics. The simplest instanton is given by the lower half of a four-sphere ($S^4$) glued at its equator to the waist of a four-dimensional de Sitter hyperboloid (dS$_4$). The metrics are given by the line elements
	\begin{subequations}
		\begin{align}
			\dd s^2_{S^4}&=\dd\tau^2+\frac{\sin[2](H\tau)}{H^2}\dd\Omega_{(3)}^2\,,\\
			\dd s^2_{\textrm{dS}_4}&=-\dd t^2+\frac{\cosh[2](Ht)}{H^2}\dd\Omega_{(3)}^2\,,
		\end{align}
	\end{subequations}
	where $\dd\Omega^2_{(3)}$ is the metric on a three-sphere of constant radius $1/H$, and the two metrics are related to each other through the following Wick rotation:
	\begin{equation}
		t= - i\left(\tau-\frac{\pi}{2H}\right)\,.
	\end{equation}
	If we decide to glue onto the `South Pole' of the half four-sphere an entire new four-sphere, we find a new instanton solution. We can actually add $n$ such four-spheres below our original instanton to create $n$ new instanton solutions. An example of this is given in figure~\ref{fig:no-boundaryManySpheres} with two additional spheres. These instanton geometries provide a large contribution to the wave function, 
	\begin{equation}
		\Psi\sim\exp[-\left(n+\frac{1}{2}\right)\frac{\mathcal{I}(S^4)}{\hbar}+i\varphi]=\exp[+\frac{4\pi^2}{\hbar}\left(\frac{2n+1}{H^2}+iH\left(q_1-\frac{1}{H^{2}}\right)^{3/2}\right)]\,,
	\end{equation}
	where $\mathcal{I}(S^4)$ is the on-shell Euclidean Einstein-Hilbert (EH) action of the four-sphere, $\varphi$ is the phase given by the de Sitter hyperboloid, and $q_1$ is the value of the scale factor squared evaluated on the final hypersurface.
	We see that the amplitude of the wave function grows exponentially with $n$, which would imply that the more four-spheres we glue below the solution, the more probable it becomes, and the wave function diverges as $n \to \infty$. This is a nonsensical result, and we would not want the path integral to sum over these geometries.
	We will see that the criterion on complex metrics studied here renders non-allowable the addition of any number of Euclidean four-spheres to the original no-boundary proposal, irrespective of the time path chosen, thus resolving the puzzle just presented.
	
	It is important to stress that the exact time path followed in the example given above is not necessarily rigid: if the boundary conditions remain respected, changes in the complex time path do not modify the action and therefore the path integral.\footnote{This is essentially due to the fundamental theorem of calculus for complex analytic functions, which implies independence of path. Therefore, the statement applies as long as the complex time paths lie in a simply connected open domain of $\mathbb{C}$ and as long as the Lagrangian for the given model is analytic (or equivalently holomorphic) on that domain, i.e., Cauchy's integral theorem must hold. As such, this cannot hold if singularities exist within the given domain. This caveat is relevant in situations where large anisotropies are present, see \cite{Bramberger:2017rbv}. In the present paper we restrict our analysis to isotropic cosmologies.} For instance, while the transition from Euclidean to Lorentzian time is instantaneous in the description given above of the original no-boundary proposal (and as depicted in the left-hand side of figure~\ref{fig:no-boundaryManySpheres}), it need not be infinitely sharp. The transition can certainly be smooth as depicted on the right-hand side of figure~\ref{fig:no-boundaryManySpheres}, and this yields the same wave function -- therefore, we consider the time paths equivalent. Moreover, the path could certainly avoid the zeroes of the scale factor where the spheres' poles meet by doing an infinitesimal deformation into the Lorentzian direction -- this is depicted by the dotted curves in the right plot of figure~\ref{fig:no-boundaryManySpheres}. Generally, such deformations would take the form
	\begin{equation}
		\dd s^2=\tau'(u)^2\dd u^2+\frac{\sin^2\left[H\tau(u)\right]}{H^2}\dd\Omega_{(3)}^2\,,
	\end{equation}
	where $u$ is a real variable with domain $\mathcal{U}\subseteq\mathbb{R}$ (usually some closed interval) parameterising the deformation $\tau(u):\mathcal{U}\mapsto\mathbb{C}$. This function is assumed to be continuously differentiable, and its derivative is denoted by a prime.
	The resulting time path is also assumed to be simple, such that it admits no closed time loop.

	\subsection{The LSKSW allowability criterion}
	
	Now that we have motivated the necessity of distinguishing between `good' and `bad' complex metrics in the context of quantum cosmology, we will present one possible criterion for making this distinction, namely the LSKSW criterion.
	
	The genesis of this criterion goes back to Louko \& Sorkin \cite{Louko:1995jw}, where they formulated a first version for scalar fields in $1+1$ dimensions. Their point was that even though the variational principle for deriving equations of motions from the EH action works with smooth and invertible metrics, non-invertible metrics are sometimes necessary in cases of physical interest, such as topology changes in two spacetime dimensions, which are necessary, e.g., to compute the fundamental vertex of string theory (`crotch' singularity). They hence desired to extend the class of metrics entering the definition of the EH action by relaxing the non-invertibility assumption, rendering topology change kinematically possible. To this aim, they studied $2$-dimensional Lorentzian analytic metrics vanishing at one single point and showed that those spacetimes can be regularised by a simple $+i\epsilon$ prescription. Crucially, they fix the sign of the regulator by requiring that the action for a massless scalar field, $S[\phi]=-\frac{1}{2}\int\dd^2x\,\sqrt{-g}\,g^{\mu\nu}\partial_\mu\phi\partial_\nu\phi$, where $g\equiv\mathrm{det}(g_{\mu\nu})$, has a positive imaginary part, $\Im(S[\phi])>0$, assuming that $\phi$ is real valued on the regularised spacetime. That way, the path integral $\int\mathcal{D}\phi\,\exp(\frac{i}{\hbar}S[\phi])$ is convergent on this regularised spacetime. Louko \& Sorkin also noticed that by diagonalising the metric, $g_{\mu\nu}=\mathrm{diag}(\lambda_{(0)},\lambda_{(1)})$,
	one could rewrite the convergence condition as $\Im[\sqrt{-g}]=\Im[\sqrt{-\lambda_{(0)}\lambda_{(1)}}]<0$, and they speculated on the possibility of defining a similar criterion for other types of matter fields such as an electromagnetic field strength.
	
	The second contribution to this criterion came from Kontsevich \& Segal \cite{Kontsevich:2021dmb}, who in the spirit of replacing the axioms of quantum field theory (QFT) characterised complex background geometries as allowable if any QFT could be consistently defined on these backgrounds. More precisely, on a spacetime manifold $\mathcal{M}$ of dimension $D\geq 2$, the complex invertible metric $\bm{g}$ is allowable if the (kinetic part of the) Euclidean action for any real non-zero $p$-form gauge field $\bm{A}$ with associated $q$-form field strength $\bm{F}=\mathbf{d}\bm{A}$, $q=p+1$, has a positive real part:
	\begin{align}
		&\mathcal{I}_q[\bm{A}]=\frac{1}{2q!}\int_\mathcal{M}\dd^Dx\,\sqrt{\det(\bm{g})}\,g^{\mu_1\nu_1}\cdots g^{\mu_{q}\nu_{q}}F_{\mu_1\cdots \mu_{q}}F_{\nu_1\cdots \nu_{q}}\,;\nonumber\\
		&\bm{g}\ \textrm{is~allowable~if}\ \forall\,q\leq D\,,\  \Re\left(\mathcal{I}_{q}[\bm{A}]\right)>0\,.\label{eq:allowable_metrics}
	\end{align}
	When $\Re\left(\mathcal{I}_{q}[\bm{A}]\right)>0$, the Euclidean path integral constructed from this action, of the form $\int\mathcal{D}\bm{A}\,\exp\left(-\mathcal{I}_{q}[\bm{A}]/\hbar\right)$, is convergent. When $q=0$, there is no interpretation in terms of the field strength of a gauge field, but this case corresponds to a scalar field mass term, as already discussed by Louko \& Sorkin \cite{Louko:1995jw} in two dimensions, and we see that their criterion matches the general proposition given here.
	As an illustrative example, the action of a real massive scalar field on a Lorentzian background reads
	\begin{align}
		S[\phi]&=-\frac{1}{2}\int\dd t\,\dd^{D-1}x\,\sqrt{-g_\mathrm{L}}\left(g_\mathrm{L}^{\mu\nu}\partial_\mu\phi\partial_\nu\phi+m^2\phi^2\right)\nonumber\\
		&=+\frac{i}{2}\int\dd\tau\,\dd^{D-1}x\,\sqrt{g_\mathrm{E}}\left(g_\mathrm{E}^{\mu\nu}\partial_\mu\phi\partial_\nu\phi+m^2\phi^2\right)=i\mathcal{I}[\phi]\,,
	\end{align}
	where we used the Wick rotation $t=-i\tau$ for the second equality to transform the Lorentzian background metric to the Euclidean one. We recover the correct relation between the Lorentzian and Euclidean actions, $iS[\phi]=-\mathcal{I}[\phi]$, and therefore, convergence of the Lorentzian path integral, which requires
	\begin{equation}
		\Im\left(S[\phi]\right)>0\ \Leftrightarrow\ \Im\left[\sqrt{-g_\mathrm{L}}\right]<0\,,\label{eq:convL}
	\end{equation}
	is completely equivalent to convergence of the Euclidean path integral, which requires
	\begin{equation}
		\Im\left(i\mathcal{I}[\phi]\right)=\Re\left(\mathcal{I}[\phi]\right)>0\ \Leftrightarrow\ \Re\left[\sqrt{g_\mathrm{E}}\right]>0\,.\label{eq:convE}
	\end{equation}
	In what follows (in particular, for the derivation of a pointwise criterion in the next paragraph), it will be less ambiguous to work with the Euclidean version of the criterion, as stipulated in \eqref{eq:allowable_metrics}. Nevertheless, it is important to understand that the criterion really applies to any complex metrics, whether Euclidean, Lorentzian, or anything in between. Therefore, in the example above, the requirements $\Re[\sqrt{g}]>0$ and $\Im[\sqrt{-g}]<0$ should be completely equivalent.
	Note, however, that this equivalence only holds when a particular choice of sign is taken for the square root of the determinant of the metric, namely $\sqrt{-g}=-i\sqrt{g}$. This choice is consistent with the $+$ sign usually taken when evaluating the square root of the determinant of a Lorentzian metric.

	As proven in \cite{Kontsevich:2021dmb} and also rederived in \cite{Witten:2021nzp,Lehners:2021mah}, the definition of allowable metrics \eqref{eq:allowable_metrics} can be re-expressed as a very useful pointwise criterion: at a point $x\in\mathcal{M}$, the metric can be put in diagonal form\footnote{The diagonalisation should be done such that the coordinates remain real, i.e., it is only the matrix coefficients that may be complex valued.} $g_{\mu\nu}(x)=\lambda_{(\mu)}(x)\delta_{\mu\nu}$; then the metric is allowable if and only if
	\begin{equation}
		\Sigma(x)\equiv\sum_{\mu=0}^{D-1}\left|\mathrm{Arg}\left[\lambda_{(\mu)}(x)\right]\right|<\pi\quad\forall\ x\in \mathcal{M}\,.\label{eq:pointwise_criterion}
	\end{equation}
	This is found by considering the pointwise inequality
	\begin{equation}
		\Re[\sqrt{g}\,g^{\mu_1\nu_1}\cdots g^{\mu_q\nu_q}F_{\mu_1\cdots\mu_q}F_{\nu_1\cdots\nu_q}]>0\,,
	\end{equation}
	which follows from \eqref{eq:allowable_metrics}, for successive values of $q$. This is the version of the criterion we will use in the rest of this paper. We will refer to $\Sigma(x)$ as the LSKSW function and to the inequality $\Sigma(x)<\pi$ as the LSKSW or allowability bound.
	
	A few comments before going further. Kontsevich \& Segal require working with \emph{real} gauge fields because they aim to define quantum states in Hilbert space using path integrals. A Euclidean transition amplitude, written in the Schr\"odinger picture as a bra and a ket sandwiching an evolution operator, $\Psi[1\to2]=\bra{2}\exp\big(-\tau\hat H/\hbar\big)\ket{1}$ (where $\hat H$ is the Hamiltonian operator), corresponds to a path integral fixed at its two ends. In this sense, a quantum state composed of an evolved ket $\ket{\psi_1}=\exp\big(-\tau\hat H/\hbar\big)\ket{1}$ can be viewed as a path integral fixed at only one of its ends. Then, if the matter fields are real, the path integral over them can be carried out along the real line and integration contours can be defined locally in field space. However, if the matter fields are allowed to take complex values, then the path integral is defined via analytic continuation, and the thimbles in the complexified field space, along which one integrates, depend on the two ends of integration. Thus, these thimbles cannot be specified {\it a priori} when only one end is fixed, obstructing the definition of quantum states. One might wonder if this argument is too strong and if well-defined transition amplitudes might be enough. Kontsevich \& Segal use their axioms to argue that they lead to unitary QFTs, which is evidently a desirable property. It is not clear at present if such a proof could be extended to complexified fields (presumably with additional restrictions). Another argument for why one should only consider real matter fields is that for $p$-forms with $p \geq 1$ (i.e., not scalar fields) the field configurations also include some integer variables often called the monopole numbers, the instanton numbers, or the fluxes, and integers cannot be analytically continued, though, as noted in \cite{Witten:2021nzp}, this would only lead to a weaker global, and not a local, restriction. Finally, we point out that the addition of a complex scalar field in the no-boundary proposal introduces strong non-localities since in that case the South Pole values of the scalar field are found to depend on the final size of the universe \cite{Jonas:2021ucu}, leading to the highly non-local picture of a universe constantly `recalculating itself' as it expands -- this effect can be avoided if the scalar field remains real and at an extremum of the potential.
	
	The final step in this allowability criterion, taken by Witten in \cite{Witten:2021nzp}, consists in applying the criterion \eqref{eq:allowable_metrics} not only to QFTs on curved backgrounds, but also to quantum gravity. This application is speculative because here no formal proof exists (simply because quantum gravity is not a complete theory). Nevertheless, the criterion is supported at the semi-classical level by the fact that all `good' examples of complex metrics leading to sensible results were found to be allowable, while all `bad' examples were non-allowable. This is also the case for the pathological metric described in the previous subsection, as we will show in section \ref{sec:quantumBounces}.
	
	Here we apply the allowability criterion to quantum cosmology, including off-shell configurations, significantly extending what was initiated in ref.~\cite{Lehners:2021mah}. For early universe cosmological models, there is no doubt that such a criterion must exist, and the LSKSW criterion is the only sensible one we have at our disposal for now. Given the support we find for this criterion, we suspect that it will eventually be derived directly from some underlying physical principles of quantum gravity, though the full quantum gravity criterion might well end up being even more restrictive.

	\subsection{Computational methods to determine allowability}\label{subsec:methods}
	
	The condition \eqref{eq:pointwise_criterion} provides a conceptually simple way of characterising allowable complex metrics. In practice, determining whether a given metric satisfies this condition or not can, however, be a rather involved task. To simplify it, we have devised several methods, which are particularly well suited to minisuperspace models in which the metric is parameterised by several functions of time only. The first two methods are conceptually simple, but not always conclusive, while the method presented in section \ref{subsec:minmaxcurves} always works as a criterion of allowability.

	\subsubsection{Optimising deformations of the time contour}
	
	In \cite{Witten:2021nzp,Lehners:2021mah}, the on-shell metric of the no-boundary solution was found to be allowable by deforming the time contour in the complex plane so as to find a time path where the LSKSW bound was satisfied everywhere. Similarly, here we consider a metric to be allowable if a continuous and differentiable deformation of the time contour, keeping boundary points fixed, can render it allowable. As already explained, this is justified by the fact that such a change does not alter the value of the action (as long as there are no holes in the domain), nor any observable quantities such as the transition amplitude calculated from the path integral.
	
	In general, the aim is, therefore, to try to find one time path that satisfies the LSKSW bound. A time-consuming (and not very efficient) way is to run an optimisation algorithm on the value of the LSKSW function, where the optimisation parameters characterise the time path in a judicious way, until finding a set of parameters for which $\Sigma(t)<\pi\ \forall\,t$.\footnote{We work with homogeneous metrics for the rest of this work, hence the LSKSW function will only depend on time, not on space.}
	
	It is useful to work with a normalised time variable that is such that the boundary conditions of a particular solution are set at an initial time $t=0$ and a final time $t=1$. The idea is then to promote the real, straight path from $t=0$ to $t=1$ to a complex path $t(u):[0,1]\mapsto\mathbb{C}$ such that $t(0)=0$ and $t(1)=1$. The function $t(u)$ may depend on a number of real parameters $c_1$, $c_2$, ... (say $n$ of them, collectively written $\vec{c}\,$), so let us actually denote the parameterised path by $t(u;\vec{c}\,)$. Given some compact domain $\mathcal{A}\subset\mathbb{R}^n$ for the set of parameters, the optimisation procedure then corresponds to finding whether
	\begin{equation}
		\min_{\vec{c}\,\in\mathcal{A}}\max_{u\in[0,1]}\Sigma(t(u;\vec{c}\,))<\pi\,.\label{eq:opt}
	\end{equation}
	This equation uses the fact that demanding $\Sigma(t(u;\vec{c}\,))<\pi\ \forall\,u\in[0,1]$ at fixed $\vec{c}\,$ is equivalent to requiring the maximal value of $\Sigma(t(u;\vec{c}\,))$ for $u$ in the range from $0$ to $1$ to lie below $\pi$ (again at fixed $\vec{c}\,$). Then, this becomes a minimisation problem in the parameter space $\mathcal{A}$, where one succeeds as soon as a set of parameters $\vec{c}\,$ is found such that the LSKSW bound is satisfied.\footnote{A word on methodology: the numerical implementation of the above optimisation problem has been carried out using the basin-hopping algorithm (see, e.g., \cite{1998cond.mat..3344W,Olson:2012}) of the \texttt{SciPy Python} package \cite{Virtanen:2019joe}. The procedure is set up such that there are two nested basin-hopping optimisations: at every point $\vec{c}$ in the parameter space $\mathcal{A}$ explored by the outer minimisation, an inner maximisation is called over $u\in[0,1]$ to find the global maximum of $\Sigma$. If this maximum is below $\pi$, the algorithm is stopped; otherwise, the optimisation over $\mathcal{A}$ is continued until a certain number of iterations has been reached. This cannot guarantee that the value on the left-hand side of \eqref{eq:opt} is truly a global minimum, but it is often sufficient to find local minima. When performing an optimisation for many different metrics, we took advantage of parallelisation using the \texttt{multiprocessing} package.} If the global minimum does not respect the LSKSW bound, then this means that either the parameterisation of the time path is not good enough, or the bound is actually impossible to satisfy (in which case the metric is non-allowable). However, this method cannot be conclusive in such a case, as it would require computationally checking an infinite number of possibilities. We will present more conclusive techniques in the next sub-sections, but this numerical optimisation is useful to develop intuition about allowable metrics. For this purpose, we have used the following parameterisation (with a $5$-dimensional parameter space), which is a linear combination of the straight path with the addition of the first three terms of a Fourier series in the complex time direction, together with a power-law function (inspired by the path that worked in \cite{Lehners:2021mah}) that starts by following the direction in which the metric is Euclidean and then smoothly interpolates to the direction in which it is Lorentzian: 
	\begin{align}
		t(u;c_1,\dots,c_5)=&~c_5\big[u+ic_1\sin(\pi u)+ic_2\sin(2\pi u)+ic_3\sin(3\pi u)\big]\nonumber\\
		&+(1-c_5)e^{i(\vartheta-\pi/2)}   \Big[\left(1-(1-u)^{c_4}\right)\sin(\vartheta)+iu^{c_4}\cos(\vartheta)\Big]\,.
	\end{align}
	In the above, $\vartheta\equiv\mathrm{Arg}(-N)$ is used to appropriately rotate the power-law part of the path as a function of the fixed lapse $N$ for a given (on- or off-shell) solution. Various numerical ranges have been explored for $c_1,\ldots,c_5$, but as an example, many successful paths could be found with the Fourier coefficients $c_1,c_2,c_3\in[-1/2,1/2]$, power-law coefficient $c_4\in[1,10]$, and linear weighting coefficient $c_5\in[0,1]$. In the figures presented below, this numerical optimisation (when successful) was used in conjunction with the more analytic techniques that we will describe next.

	\subsubsection{Checking for uncrossable ridges}
	
	There is a neat verification that can be performed in order to check whether a certain solution may immediately violate the LSKSW bound for any time paths.
	To do so, let us split our diagonalised homogeneous metric into temporal and spatial parts. For instance, we can write quite generally our metric ansatz in the form
	\begin{equation}
		\dd s^2=-f(t)\dd t^2+\sum_{j=1}^{D-1}q_{(j)}(t)(\sigma^j)^2=-f(t(u))t'(u)^2\dd u^2+\sum_{j=1}^{D-1}q_{(j)}(t(u))(\sigma^j)^2\,,
		\label{eq:homoisometric}
	\end{equation}
	where the $\sigma^j$s are differential 1-forms on the spatial hypersurface,
	and then the LSKSW function reads
	\begin{equation}
		\Sigma(t(u))\equiv\Sigma_\textrm{temporal}(t(u))+\Sigma_\textrm{spatial}(t(u))=\left|\mathrm{Arg}\left[-f(t(u))t'(u)^2\right]\right|+\sum_{j=1}^{D-1}\left|\mathrm{Arg}\left[q_{(j)}(t(u))\right]\right|\,.\label{eq:Sigmasplit}
	\end{equation}
	Due to the absolute values, one has $\Sigma_\textrm{temporal}(t(u))\geq 0$ and $\Sigma_\textrm{spatial}(t(u))\geq 0$, thus one always has $\Sigma(t(u))\geq\Sigma_\textrm{temporal}(t(u))$ and $\Sigma(t(u))\geq\Sigma_\textrm{spatial}(t(u))$.
	From this, the existence of a connected path from $t(0)=0$ to $t(1)=1$ where $\Sigma_\textrm{spatial}(t(u))<\pi\ \forall\,u\in[0,1]$ is a necessary (but not sufficient) condition for the existence of a path satisfying the LSKSW bound.
	Thus, if there is an obstruction to the existence of any such connected paths, it is a sufficient criterion to declare that the LSKSW bound is violated (or at best saturated, i.e., with $\Sigma=\pi$) for this metric. The advantage of this criterion is that it only depends on the position $t(u)$ of the time paths in the complex plane and not on the tangent $t'(u)$ of the paths.
	
	In practice, we must typically restricts ourselves to a bounded, closed domain of the complex plane (therefore compact; let us call it $\mathcal{B}\subset\mathbb{C}$). If the real part of the time path grows monotonically (so that Lorentzian time never stops or turns around), then this sets the restriction $\Re(t)\in[0,1]$. Similarly, we can set some bound regarding how far in the imaginary direction a time path can go, e.g., $\Im(t)\in[-1,1]$ (in this case, $\mathcal{B}$ is rectangular).
	We then simply need to check whether there exists a simply connected subset $\mathcal{S}\subseteq\mathcal{B}$ where $\Sigma_\textrm{spatial}(t)<\pi\ \forall\,t\in\mathcal{S}$, which contains the start and end points $t=0$ and $t=1$. If there is no such region $\mathcal{S}$, then this means that time paths within $\mathcal{B}$ inevitably have to cross a region where $\Sigma_\textrm{spatial}(t)\geq\pi$, hence violating the LSKSW bound. In many cases, even if $\mathcal{B}$ is compact, this can be enough to convince oneself that the result in fact holds true for $\mathcal{B}$ extending to positive and negative imaginary infinity.
	
	\begin{figure}[t]
		\centering
		\includegraphics[width=0.4\textwidth]{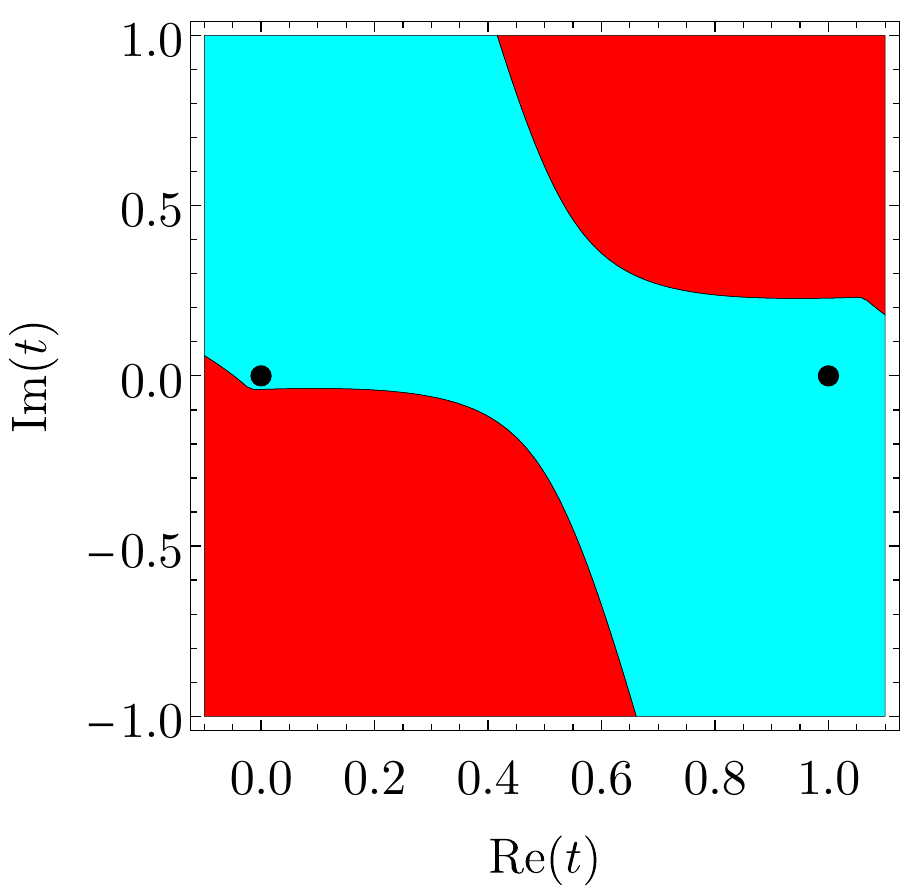}\hspace*{1cm}
		\includegraphics[width=0.4\textwidth]{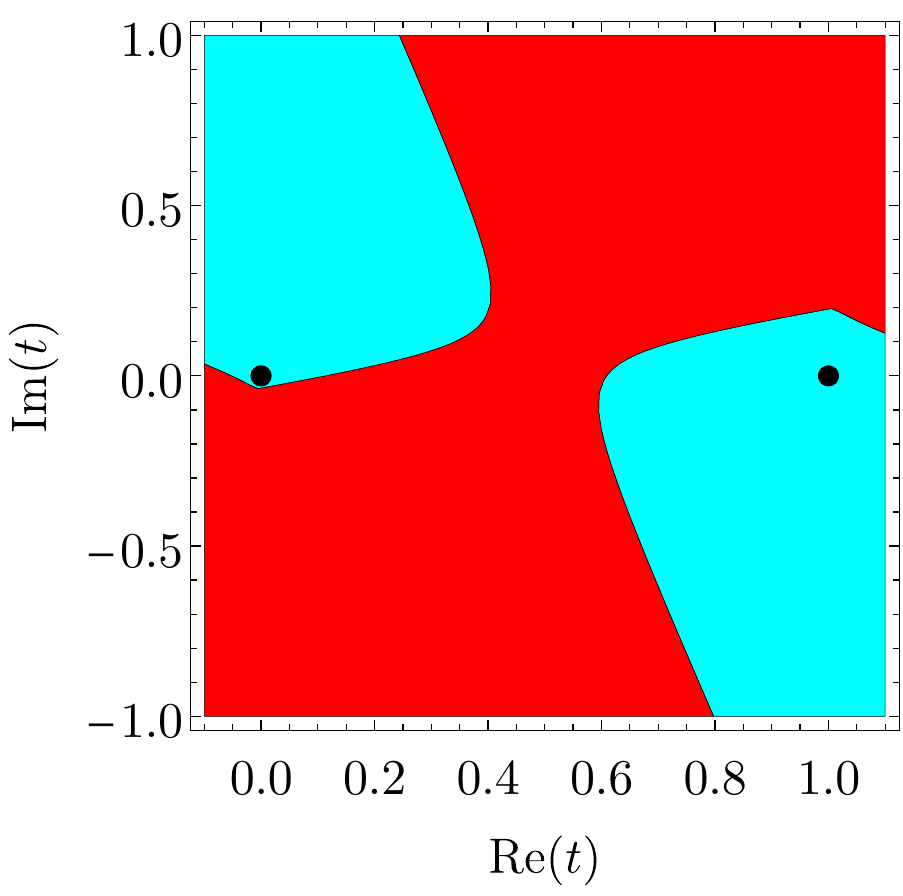}
		\caption{An example of the ridge obstruction picture in the complexified $t$ plane.
			The `seas' (cyan regions) are where $\Sigma_\textrm{spatial}(t)<\pi$, and the `ridges' (red regions) are where $\Sigma_\textrm{spatial}(t)\geq\pi$. On the left,
			one has an example where the ridges do not present an obstruction and where we can find a simply connected smooth path between $t=0$ and $t=1$ (represented in both pictures by the black dots) along which $\Sigma_\textrm{spatial}(t)<\pi$. Whether the LSKSW bound can actually be satisfied along such a path is however not guaranteed and must be decided based on other methods. On the right,
			one has an example where there is a ridge obstructing the way between $t=0$ and $t=1$. This implies that the LSKSW bound is necessarily violated.}
		\label{fig:ridge_example}
	\end{figure}
	
\begin{figure}[t]
		\centering
		\includegraphics[width=0.4\textwidth]{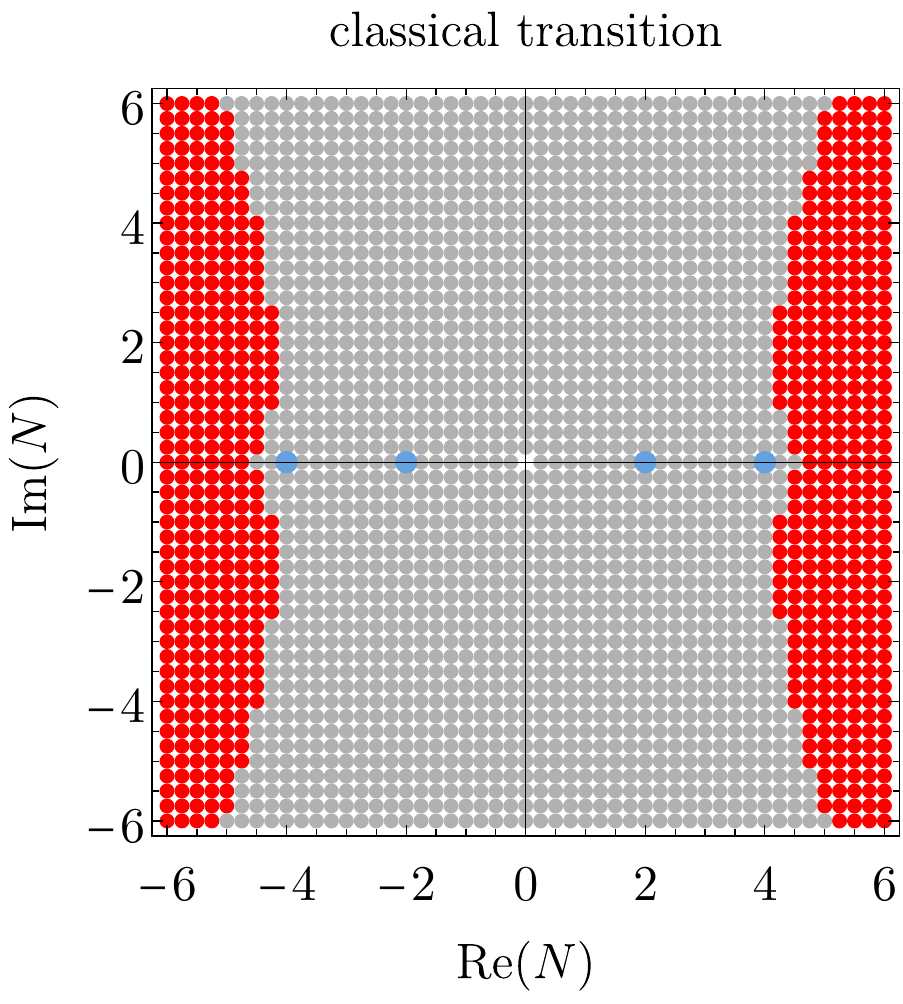}
		\hspace*{1cm}
		\includegraphics[width=0.4\textwidth]{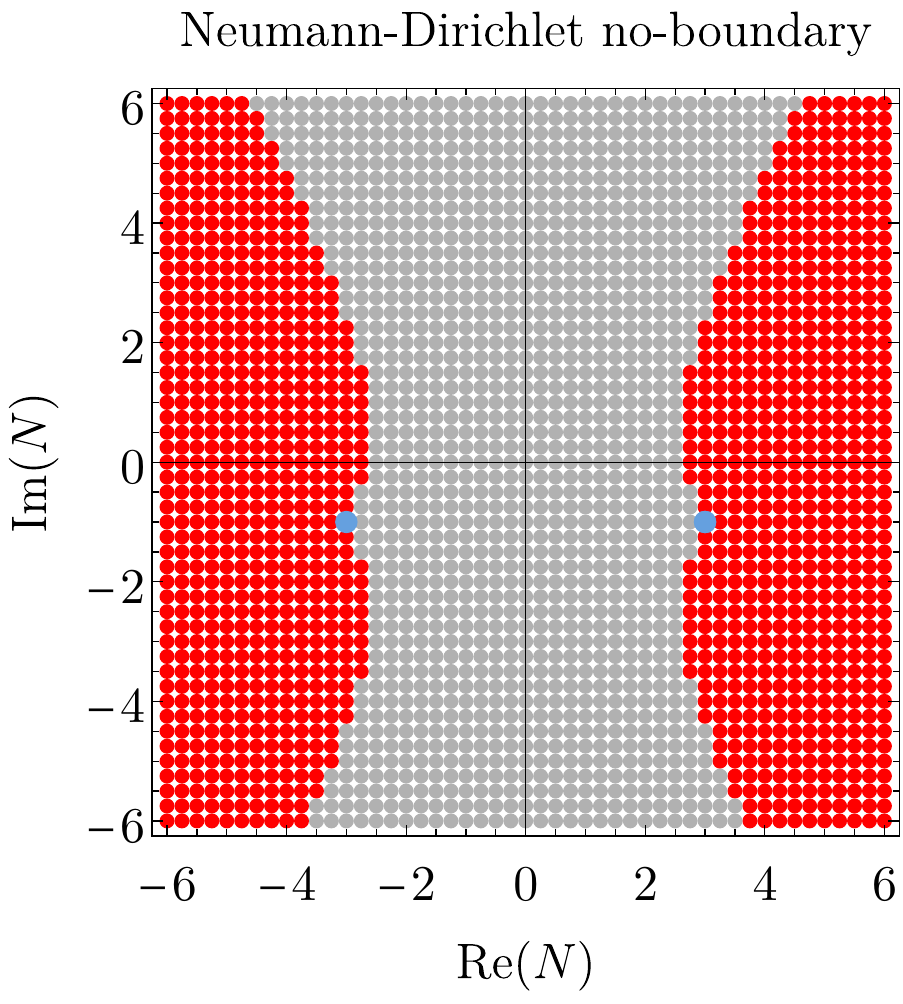}
		\includegraphics[width=0.4\textwidth]{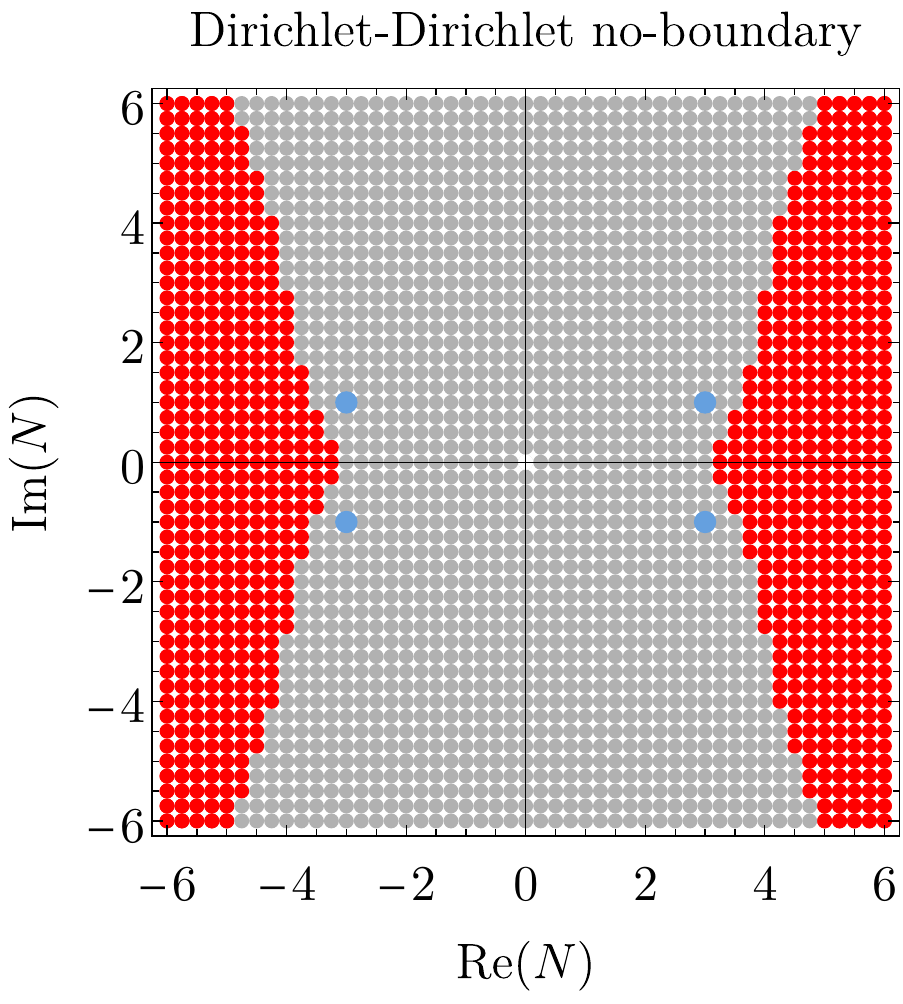}
		\caption{Results of the walking algorithm for the different models that will be studied in this work. In red: points where a ridge is found (the LSKSW bound is violated). In grey: points where no obstruction is found, hence where further investigation is needed. Saddle points (i.e., classical solutions) are represented in blue.}
		\label{fig:walkingalgo_results}
	\end{figure}

	Pictorially, such an obstruction resembles a `ridge' exceeding the `sea' of $\Sigma_\textrm{spatial}<\pi$ points and blocking the way between $t=0$ and $t=1$; as an example see figure~\ref{fig:ridge_example}.
	This `no-ridge criterion' is very useful when we start investigating specific early universe cosmological models.
	Indeed, we will be concerned with homogeneous and isotropic metrics, which are solutions to general relativity with a cosmological constant for various boundary conditions. In particular, we will study classical transitions, as well as the no-boundary proposal with Neumann-Dirichlet and Dirichlet-Dirichlet boundary conditions.
	The path integral for all those scale factor solutions will involve summing over geometries in the complex plane for the lapse $N$. Thus, for any fixed value of $N$, we have a metric for which we can write down the LSKSW function and apply the no-ridge criterion. The way we assess this criterion is by using a `walking algorithm': upon discretising the region $\mathcal{B}$, one starts from $t=0$ and explores the complex time plane by trying to go forwards in $\Re(t)$ in small increments and reach $t=1$, while always staying `in the water', i.e., keeping $\Sigma_\textrm{spatial}(t)<\pi$. If it is impossible, we declare the metric non-allowable. While this is approximate in the sense that the numerical procedure has limited precision (given by the inverse lattice spacing) and in the sense that it is restricted to a bounded region of the complex plane, we have found the method to be perfectly robust. Some results are displayed in figure~\ref{fig:walkingalgo_results}. It is important to notice that the no-ridge criterion already renders non-allowable large portions of the geometries, especially at large $|N|$. This indicates the potential power of the LSKSW criterion to constrain complex metrics that should be non-allowable in path integrals.
	
	\subsubsection{Allowable time domains}
	\label{subsec:minmaxcurves}
	
	We are now in a position to describe our most powerful method to determine allowability. To explain it, let us turn our attention to a particular homogeneous and isotropic metric of the form of eq.~\eqref{eq:homoisometric}, which is of common use in minisuperspace quantum cosmology \cite{Halliwell:1988ik}, and which we use repeatedly throughout this work:
	\begin{equation}
		\dd s^2=-\frac{N^2}{q(t)}\dd t^2+q(t)\dd\Omega_{(D-1)}^2=-\frac{N^2}{q(t(u))}t'(u)^2\dd u^2+q(t(u))\dd\Omega_{(D-1)}^2\,.
	\end{equation}
	For this metric, the LSKSW criterion [where the LSKSW function is of the form of \eqref{eq:Sigmasplit}] reads
	\begin{equation}
		\Sigma(t(u))=\left|\mathrm{Arg}\left[-\frac{N^2}{q(t(u))}t'(u)^2\right]\right|+(D-1)\left|\mathrm{Arg}\big[q(t(u))\big]\right|<\pi\quad\forall\,u\in[0,1]\,.\label{eq:LSKSWmini}
	\end{equation}
	Suppose we are dealing with a geometry that passed the no-ridge criterion, i.e., for which there exist time paths such that $\Sigma_\textrm{spatial}(t(u))=(D-1)|\mathrm{Arg}[q(t(u))]|<\pi\ \forall\,u\in[0,1]$; this tells us that on these paths the quantity $\pi-(D-1)|\mathrm{Arg}[q(t(u))]|$ is always positive.
	Then, our goal is to isolate $t'(u)$ from the temporal part of the LSKSW function in order to obtain an ordinary differential inequality -- said differently, at each point in the complexified time plane there will be a range of directions $t'(u)$ which are such that the LSKSW bound remains satisfied. The inequality \eqref{eq:LSKSWmini} is equivalent to (unless otherwise stated, the equations must hold for any values of $u$ in the range $[0,1]$)
	\begin{equation}
		-\pi+(D-1)\left|\mathrm{Arg}\big[q(t(u))\big]\right|<\mathrm{Arg}\left[-\frac{N^2}{q(t(u))}t'(u)^2\right]<\pi-(D-1)\left|\mathrm{Arg}\big[q(t(u))\big]\right|\,.
	\end{equation}
	Upon using properties of the complex argument, this condition can be rewritten as
	\begin{equation}
		F_\mathrm{min}(t(u))<\mathrm{Arg}\left[t'(u)\right]<F_\mathrm{max}(t(u))\,,\label{eq:argtpineq}
	\end{equation}
	where we defined
	\begin{align}
		F_\mathrm{min}(t(u))&\equiv-(n+1)\pi+\frac{D-1}{2}\left|\mathrm{Arg}\big[q(t(u))\big]\right|-\frac{1}{2}\mathrm{Arg}\left[\frac{N^2}{q(t(u))}\right]\,,\nonumber\\
		F_\mathrm{max}(t(u))&\equiv-n\pi-\frac{D-1}{2}\left|\mathrm{Arg}\big[q(t(u))\big]\right|-\frac{1}{2}\mathrm{Arg}\left[\frac{N^2}{q(t(u))}\right]\,,
	\end{align}
	and where $n\in\{-2,-1,0,1\}$ appears due to the fact that the principal value of the argument of a complex number has to remain within the range $(-\pi,\pi]$.
	
	The inequalities \eqref{eq:argtpineq} imply that, at any point $t(u)\in\mathbb{C}$, the slope of a time path passing through that point has to lie within some interval delimited by $F_\mathrm{min}(t(u))$ and $F_\mathrm{max}(t(u))$ in order to respect the LSKSW bound at that point. Therefore, one can actually find lower and upper bounds on the allowable time paths. Indeed, since the paths all have to start at $t(0)=0$, one can solve the corresponding initial value problems (IVPs) for the ordinary differential equations\footnote{Since $F_\mathrm{min}$ and $F_\mathrm{max}$ can take four different values each in order to have the correct principal value for the complex argument, one actually has to solve four IVPs for $t_\mathrm{min}(u)$ and four for $t_\mathrm{max}(u)$. It is nevertheless straightforward to then verify which solutions correspond to the correct lower and upper bounds.} $\mathrm{Arg}[t_\mathrm{min}'(u)]=F_\mathrm{min}(t_\mathrm{min}(u))$ and $\mathrm{Arg}[t_\mathrm{max}'(u)]=F_\mathrm{max}(t_\mathrm{max}(u))$. Then, it follows from Petrovitsch's theorem \cite{Petrovitch:1901} that time paths $t(u)$ respecting the inequalities \eqref{eq:argtpineq} have to lie within the curves $t_\mathrm{min}(u)$ and $t_\mathrm{max}(u)$ for all values of $u>0$. This means that the LSKSW bound is satisfied if and only if the region within the curves $t_\mathrm{min}(u)$ and $t_\mathrm{max}(u)$ contains the required end point $t=1$.
	
	This method is powerful since one can use it to conclusively determine whether or not a given solution has time paths respecting the LSKSW bound, without having to find (or optimise) an actual allowable path. Practically, it often involves solving a few IVPs by resorting to numerical algorithms, which always comes with some level of numerical imprecision, but the method has proved itself to be quite robust. We will show examples of how this method is used in the subsequent sections.

	\subsection{Comment on Lorentzian metrics}\label{subsec:lorentzian_metrics}
	
	Here we show that any homogeneous Lorentzian metrics cannot be changed by a (smooth) complex deformation of the time path into geometries that respect the LSKSW bound everywhere. In fact, we show that these metrics at best saturate the LSKSW bound.
	
	We consider a homogeneous metric ansatz of the form (in particular, it could be a Bianchi spacetime of any type)
	\begin{equation}
		\dd s^2=-N^2\dd t^2+\sum_{j=1}^{D-1}a_{(j)}^2(t)(\sigma^j)^2\,,
		\label{eq:homoisometric2}
	\end{equation}
	whose LSKSW function reads
	\begin{equation}
		\Sigma(t)=\left|\mathrm{Arg}\left[-N^2\right]\right|+\sum_{j=1}^{D-1}\left|\mathrm{Arg}\left[a_{(j)}^2(t)\right]\right|\,.
	\end{equation}
	If the metric \eqref{eq:homoisometric2} is real and Lorentzian for some solutions $a_{(j)}(t)\in[a_0,a_1]\ \forall\,t\in[t_0,t_1]$, then in particular the lapse $N$ and the scale factors squared $a_{(j)}^2(t)$ are real valued everywhere, hence one immediately concludes that the LSKSW bound is saturated with $\Sigma(t)=\pi\ \forall\,t\in[t_0,t_1]$.
	
\begin{figure}[t]
		\centering
		\includegraphics[width=0.45\textwidth]{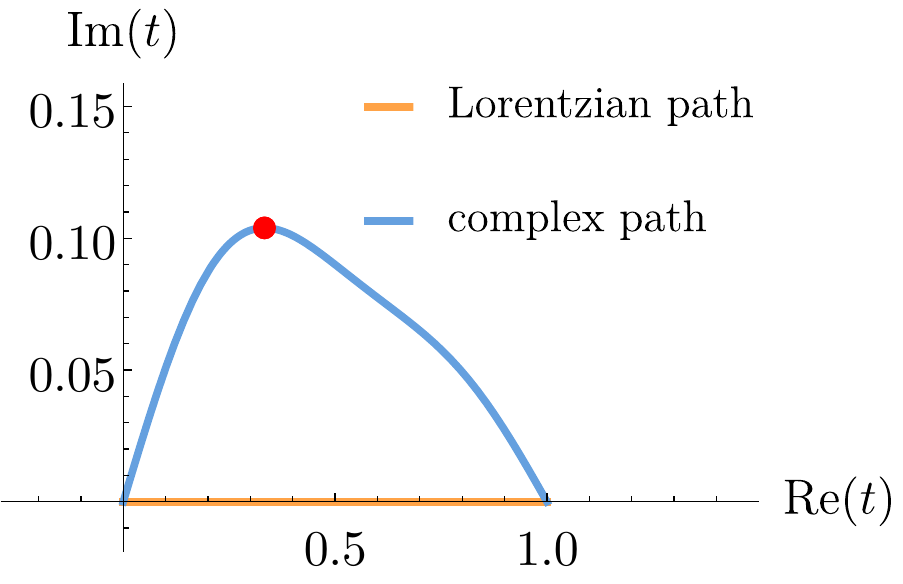}
		\caption{Example of a smooth deformation of the time path between $t_0=0$ and $t_1=1$ in the complex plane. The point $t(u_\star)$ is plotted in red, where the tangent of the complex path is parallel to the real axis, and thus where $\mathrm{Arg}[t'(u_\star)]=0$.}
		\label{fig:lorentzian_time_path}
	\end{figure}

	If we now consider complex time paths with the same boundaries $t(0)=t_0\in\mathbb{R}$, $t(1)=t_1\in\mathbb{R}$, the metric and LSKSW function respectively become
	\begin{subequations}
		\begin{align}
			\dd s^2&=-N^2t'(u)^2\dd u^2+\sum_{j=1}^{D-1}a_{(j)}^2(t(u))(\sigma^j)^2\,,\\
			\Sigma(t(u))&=\left|\mathrm{Arg}\left[-N^2t'(u)^2\right]\right|+\sum_{j=1}^{D-1}\left|\mathrm{Arg}\left[a_{(j)}^2(t(u))\right]\right|\,.
		\end{align}
	\end{subequations}
	Then for $t(u)\in\mathbb{C}\ \forall\,u\in(0,1)$, the scale factor becomes complex, and the spatial part of the LSKSW function may become non-zero. Thus, any hope of reducing $\Sigma$ below $\pi$ lies in having a time path with a non-trivial direction such that $\Sigma_\textrm{temporal}<\pi-\Sigma_\textrm{spatial}$. However, for the time path to start and end on the real-$t$ axis at $t_0$ and $t_1$, respectively, the time trajectory must `turn around' in the imaginary direction, and so there must exist at least one point $u_\star\in(0,1)$ along the trajectory where $\Im[t'(u_\star)]=0$ and so where $t'(u_\star)$ is purely real and positive (assuming again the real part of time always grows monotonically), hence where $\mathrm{Arg}[t'(u_\star)]=0$. See figure~\ref{fig:lorentzian_time_path} for a depiction. Therefore, the temporal part of the LSKSW function equates $\pi$ at that point, and thus we find that $\Sigma(t(u_\star))\geq\pi$. Consequently, the LSKSW bound is violated (or again at best saturated\footnote{Saturation is generally not even realised with a complex time path here since it could only occur if $a_{(j)}^2(t(u_\star))$ had a phase of $2n\pi$, $n\in\{1,2,\cdots\}$, in every spatial direction $j$ (a phase of $0$ is excluded since $t(u_\star)$ itself would have a non-zero phase). However, since the phases of the $a_{(j)}^2$s vary continuously from $0$ at the start of the path, one would necessarily cross another location where at least one of the phases is $\pi$, and hence where the LSKSW bound is violated.}).
	
	The way to regularise all these metrics is by adding a factor $(1\mp i \epsilon)$, $0<\epsilon\ll 1$, to the time-time component of the metric \cite{Witten:2021nzp} (see also \cite{Visser:2021ucg}). This is not a smooth deformation, in the sense that the two choices of $\mp$ sign are not smoothly related \cite{Witten:2021nzp}. This regularisation implies a change in the value of the action and hence is a true deformation of the metric and not just a rewriting. As we will see, there are contexts in which such an $i\epsilon$ deformation is useful.
	
	To end this section, let us mention that saturating the LSKSW bound (when $\Sigma=\pi$ everywhere, as in the case of real Lorentzian metrics described in this subsection) is synonym of conditionally convergent path integrals, which may or may not be problematic. Indeed, while reasonable QFTs may not be properly defined on backgrounds violating the LSKSW bound ($\Sigma>\pi$) since they lead to divergent path integrals, this is not necessarily true for metrics saturating the bound. Conditionally convergent integrals are more subtle to treat than traditional Gau\ss{}ian-like integrals, and as path integrals they dependent on
	the order of integration. Therefore, absolute convergence is favorable, and that is why the LSKSW bound ($\Re(\mathcal{I})>0$ or $\Sigma<\pi$) does not allow for equality, only strict inequality.
	Let us note that Picard-Lefschetz theory is exactly the right tool to define conditionally convergent integrals, re-writing them as sums of absolutely convergent integrals (see, e.g., \cite{Witten:2010cx,Feldbrugge:2017kzv}). Thus, one could say that Picard-Lefschetz theory provides a way of unambiguously defining what is meant by a conditionally convergent integral, removing the ambiguities with orders of integration. Applying this procedure to gravity would thus be the obvious thing to do \cite{Feldbrugge:2017kzv}, but there is a subtlety, which we will encounter in all specific examples studied below \cite{Witten:2021nzp,Lehners:2021mah}, namely that the resulting integration contours run into regions that do not satisfy the allowability bound -- thus, in all cases we have studied, we find that the steepest descent integration contours provided by Picard-Lefschetz theory are cut off.

	\section{Classical transitions}
	\label{sec:classicalTransitions}
	
	The first cosmological geometries for which we are going to study the allowability criterion are those related to classical transitions in a closed, 4-dimensional universe. Assuming general relativity with a positive cosmological constant $\Lambda$ without matter and using the $3+1$ decomposition of the metric where $N$ is the lapse, $\bm{\beta}$ is the shift, and $\bm{h}$ is the metric on spatial hypersurfaces, the transition amplitude between two three-dimensional hypersurfaces $\Sigma_{0},\Sigma_{1}\subset\mathcal{M}$ can be computed via the gravitational path integral
	\begin{equation}
		\Psi[\Sigma_0\to \Sigma_1]=\int\limits_{\Sigma_0}^{\Sigma_1}\mathcal{D}N\,\mathcal{D}\bm{\beta}\,\mathcal{D}\bm{h}\,\exp(\frac{i}{\hbar}\int\limits_\mathcal{M}\dd^4x\,\sqrt{-g}\left(\frac{R}{2}-\Lambda\right)+\sum_{n=0,1}b_n\int\limits_{\Sigma_n}\dd^3x\,\sqrt{\mathrm{det}(\bm{h})}\,K)\,,
	\end{equation}
	where $R$ is the four-dimensional Ricci scalar and $K$ is the trace of the three-dimensional extrinsic curvature tensor on the hypersurfaces $\Sigma_{0,1}$ bounding the $4$-manifold $\mathcal{M}$. The coefficients $b_n \in \{-1,0,1\}$ of the Gibbons-Hawking-York surface terms determine the type of boundary conditions one may impose. We will be interested in transitions between large, classical values of the scale factor, hence $b_0=-1$ and $b_1=1$. For such transitions, we would like to study both the saddle points and the off-shell geometries that are summed over in the path integral. 
	
	As is often the case when considering early universe cosmological models, we restrict the metric to a minisuperspace model.
	We further assume closed spatial sections, so the spacetime's manifold topology is $\mathbb{R}\times S^3$.
	If we keep the minimum possible number of degrees of freedom, we may usefully write the metric as
	\begin{equation}
		\dd s^2=-\frac{N^2}{q(t)}\dd t^2+q(t)\dd\Omega_{(3)}^2\,, \label{metric}
	\end{equation}
	where $q(t)$ is the squared-scale factor and $N$ is the lapse function. This simplified ansatz enables calculations to be performed analytically. In particular, the shift integration becomes trivial, and the action turns out to be quadratic in $q$ \cite{Halliwell:1988ik},
	\begin{equation}
		S=2\pi^2\int\dd t\,N\left(-\frac{3\dot{q}^2}{4N^2}+3-\Lambda q\right)\,,
	\end{equation}
	hence the path integral over the scale factor is a simple Gau\ss{}ian integral.
	(A dot denotes a derivative with respect to $t$.)
	Despite its simplicity, this Friedmann-Lema\^itre-Robertson-Walker (FLRW) form of the metric is a good approximation of our early universe, which is experimentally known to be extremely homogeneous and isotropic (e.g., from observations of the cosmic microwave background \cite{Planck:2018nkj}).
	
	The analysis of the path integral proceeds as follows \cite{Halliwell:1988ik}: from the action, one performs the variational principle with respect to $N$ and $q$; the classical solution $\bar q$ to the resulting equation of motion for $q$, which respects the boundary conditions $q(0)=q_0$ and $q(1)=q_1$ ($q_0\leq q_1$), is then found to be
	\begin{equation}
		\bar{q}(t)=\frac{\Lambda}{3}N^2t(t-1)+\left(q_1-q_0\right)t+q_0\,.\label{eq:qbarct}
	\end{equation}
	This solution allows one to perform the integral over the scale factor by a change of variables \cite{Halliwell:1988ik,Feldbrugge:2017kzv}, reducing the action (now on-shell in the scale factor, but not in the lapse) to
	\begin{equation}
		S_\textrm{on-shell}(N)=2\pi^2\left[\frac{\Lambda^2}{36}N^3+\left(3-\frac{\Lambda(q_0+q_1)}{2}\right)N-\frac{3(q_1-q_0)^2}{4N}\right]\,.
	\end{equation}
	We are then left with an integral over the lapse, $\Psi \sim \int_\mathcal{C} \dd N\,\exp[\frac{i}{\hbar}S_\textrm{on-shell}(N)]$, along some contour $\mathcal{C}$ in the complex plane. This integral can be performed in the saddle point approximation; the four saddle points are found to be
	\begin{equation}
		N_\textrm{SP}=\frac{3}{\Lambda}\left(\pm\sqrt{\frac{\Lambda q_1}{3}-1}\pm\sqrt{\frac{\Lambda q_0}{3}-1}\right).\label{eq:SP_class_trans}
	\end{equation}
	These four saddle points are real valued as long as the scale factor is never smaller than the waist of the de Sitter hyperboloid, i.e., if $3/\Lambda\leq q_0\leq q_1$, hence the geometries they describe fall in the category of Lorentzian homogeneous metrics discussed in section \ref{subsec:lorentzian_metrics}, which as we have shown reside on the boundary of the domain of allowable complex metrics. The two inner saddle points (which are time reversals of one another) describe expanding de Sitter solutions, where the scale factor monotonically increases from $q_0$ to $q_1$, while the outer saddle points (again time reversals of one another) describe classical de Sitter bouncing solutions, where the scale factor first decreases from $q_0$ to the waist of the de Sitter hyperboloid ($q=3/\Lambda$) and from there increases again up to $q_1$.
	
	We aim to study the structure of this lapse integral in more detail, in particular the locations of the saddle points and the associated steepest descent contours. For this we must determine the allowability of the metrics that enter this integral, i.e., the metrics that satisfy the equation of motion for the scale factor, but which are off-shell regarding the lapse constraint. We may start by looking at large $\abs{N}$: writing $N=\abs{N}e^{i\mathrm{Arg}(N)}$, the classical solution is well approximated by
	\begin{equation}
		\bar{q}(t)\stackrel{\abs{N}\to\infty}{\simeq}\frac{\Lambda}{3}\abs{N}^2e^{2i\mathrm{Arg}(N)}t(t-1)\label{eq:qapprox}
	\end{equation}
	when $t\neq0,1$.
	Whatever simply connected complex time path $t(u)$ we take, it must necessarily cross the line $\Re[t(u)]=1/2$ if it evolves continuously from $t=0$ to $t=1$. Then at that point we can write $t=1/2+i\rho$ for some $\rho\in\mathbb{R}$, and so $t(t-1)=-1/4-\rho^2$ is a negative real number.
	Therefore, the spatial part of the LSKSW function at $t=1/2+i\rho$ reads
	\begin{align}
		\Sigma_\text{spatial}(t=1/2+i\rho)=3&\abs{\mathrm{Arg} \left[\bar q(1/2+i\rho)\right]}\nonumber\\
		\stackrel{\abs{N}\to\infty}{\simeq}&3\abs{\mathrm{Arg}\left[\left(-\frac{1}{4}-\rho^2\right)\frac{\Lambda}{3}\abs{N}^2e^{2i\mathrm{Arg}(N)}\right]}=3\abs{\pm\pi+2\mathrm{Arg}(N)}\,.
	\end{align}
	Since $\Sigma\geq\Sigma_\textrm{spatial}$, this means that for $\abs{N}\to\infty$, if
	\begin{equation}
		\abs{\pm\pi+2\mathrm{Arg}(N)}\geq\frac{\pi}{3}\,,
	\end{equation}
	the LSKSW bound is for sure violated and the metric is not allowable. We conclude that, asymptotically, the metrics in question are not allowable if
	\begin{equation}
		\mathrm{Arg}(N)\in(-\pi,-2\pi/3]\cup[-\pi/3,\pi/3]\cup[2\pi/3,\pi]\,,\label{eq:asymnonallow}
	\end{equation}
	leaving only two wedges centered on the Euclidean axis. Note that this means that, perhaps surprisingly, metrics for which the real part of the lapse is large are not allowable. This is due to the fact that when the real part of the lapse gets very large, the squared-scale factor starts diving to negative values for some interval of time (because the first, negative term in \eqref{eq:qbarct} will dominate for values of $t$ not near $0,1$), and thus the signature of the metric changes. 
	
	How large must $\abs{N}$ be for the approximation \eqref{eq:qapprox} to hold? To estimate this, let us use \eqref{eq:qbarct} to compute
	\begin{align}
		\bar{q}(1/2+i\rho)&=\frac{\Lambda\abs{N}^2e^{2i\mathrm{Arg}(N)}}{3}\left(-\frac{1}{4}-\rho^2\right)+\frac{q_1+q_0}{2}+i\rho(q_1-q_0)\,,\nonumber\\
		&=-\frac{\Lambda\abs{N}^2\left(1/4+\rho^2\right)}{3}e^{2i\mathrm{Arg}(N)}+\frac{(q_1+q_0)}{2}\sqrt{1+4\rho^2\left(\frac{q_1-q_0}{q_1+q_0}\right)^2}\,e^{i\varphi}\,,\label{eq:largeNvalid}
	\end{align}
	where $\varphi\equiv\mathrm{Arg}[(q_1+q_0)/2+i\rho(q_1-q_0)]$. The large $\abs{N}$ approximation is thus valid when the norm of the first complex number in \eqref{eq:largeNvalid} is much larger than the norm of the second one, which reduces to requiring
	\begin{equation}
		\abs{N}^2\gg\frac{6(q_1+q_0)}{\Lambda\left(1+4\rho^2\right)}\sqrt{1+4\rho^2\left(\frac{q_1-q_0}{q_1+q_0}\right)^2}\,.
	\end{equation}
	The combination of factors $[1+4\rho^2(q_1-q_0)^2(q_1+q_0)^{-2}]^{1/2}/(1+4\rho^2)$ is always contained between $0$ and $1$, so the strongest constraint arises when it is equal to 1, and in this case we find that the approximation is valid when
	\begin{equation}
		\abs{N}^2\gg\frac{6}{\Lambda}(q_1+q_0)\,.\label{eq:approxNrange}
	\end{equation}
	For example for $\Lambda=3$, $q_0=2$ and $q_1=10$, we find that the approximation is valid when $\abs{N}^2\gg 24$ (so if, e.g., $|N|\gtrsim 15.5$, then \eqref{eq:approxNrange} is respected by at least one order of magnitude).
	
	The other analytical result we can find is for the `Euclidean axis' $N=iN_\mathrm{E}$, $N_\mathrm{E}\in\mathbb{R}$. In this case, \begin{equation}
		\bar{q}(t)=-\frac{\Lambda}{3}N_\mathrm{E}^2t(t-1)+\left(q_1-q_0\right)t+q_0
	\end{equation}
	is real (and positive) when time is real, that is when the time path goes straight from $0$ to $1$ on the real-time axis. Also, the time-time component of the metric is positive definite. Thus, one finds $\mathrm{Arg}[\bar{q}(t)]=0$, and moreover $\mathrm{Arg}[-N^2/\bar{q}(t)]=\mathrm{Arg}[+N_\mathrm{E}^2/\bar{q}(t)]=0$. Consequently, the LSKSW function is constantly null on the real-time axis when the lapse is purely imaginary, and as such the LSKSW bound is satisfied on the whole Euclidean axis.
	Another interesting observation is that for large $N_\mathrm{E}$, we have $\bar{q}(t)\approx-\frac{\Lambda}{3}N_\mathrm{E}^2t(t-1)$, hence the metric becomes
	\begin{equation}
		\dd s^2=\frac{N_\mathrm{E}^2}{\bar{q}(t)}\dd t^2+\bar{q}(t)\dd\Omega^2_{(3)}\approx\frac{3}{\Lambda t(1-t)}\dd t^2+\frac{\Lambda}{3}N_\mathrm{E}^2t(1-t)\dd\Omega^2_{(3)}=\frac{3}{\Lambda}\dd T^2+\frac{\Lambda}{12}N_\mathrm{E}^2\cos^2(T)\dd\Omega^2_{(3)}\,,
	\end{equation}
	where we defined a new time coordinate $T$ according to $2t-1\equiv\sin T$, with $-\pi/2\leq T\leq\pi/2$. Thus, at very large $N_\mathrm{E}$ the metric approximately corresponds to a full (huge) 4-sphere. The metric describes a Euclidean geometry that starts out near zero (in fact at $q_0$), expands to a huge size and then shrinks again to almost zero (in fact to $q_1$).
	
\begin{figure}
		\centering
		\includegraphics[width=0.45\textwidth]{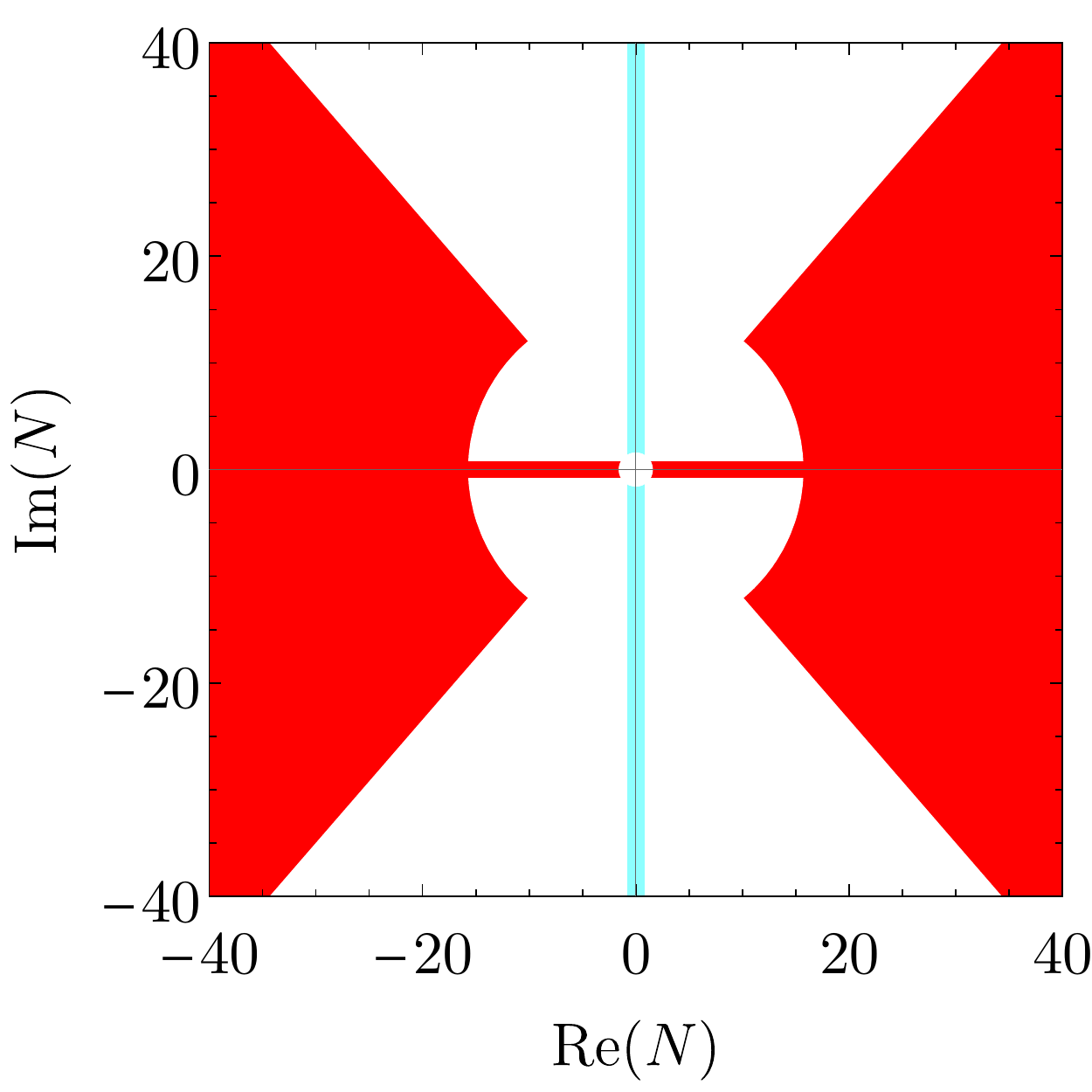}
		\caption{Analytical results for the classical transition from $q_0=2$ to $q_1=10$. The red regions represent the asymptotic region of non-allowability \eqref{eq:asymnonallow} (starting at $|N|=15.5$) and the Lorentzian axis (proved to saturate the LSKSW bound in section \ref{subsec:lorentzian_metrics}), while the Euclidean axis highlighted in cyan is allowable. The saddle points are not shown, but they are located at $N_\mathrm{SP}=\pm2,\pm4$, so the region of particular interest for the path integral (e.g., $N\lesssim 6$) remains unconstrained at this point.}
		\label{fig:asymptotic_cl_trans}
	\end{figure}

	We have shown, analytically, that $2\pi/3$ intervals for $\mathrm{Arg}(N)$ (at large $|N|$) about the Lorentzian axis always violate the LSKSW bound, while the Euclidean axis with purely imaginary lapse always respects the allowability criterion. These analytical results are summarised in figure~\ref{fig:asymptotic_cl_trans}.
	
\begin{figure}
		\centering
		\includegraphics[width=0.4\textwidth]{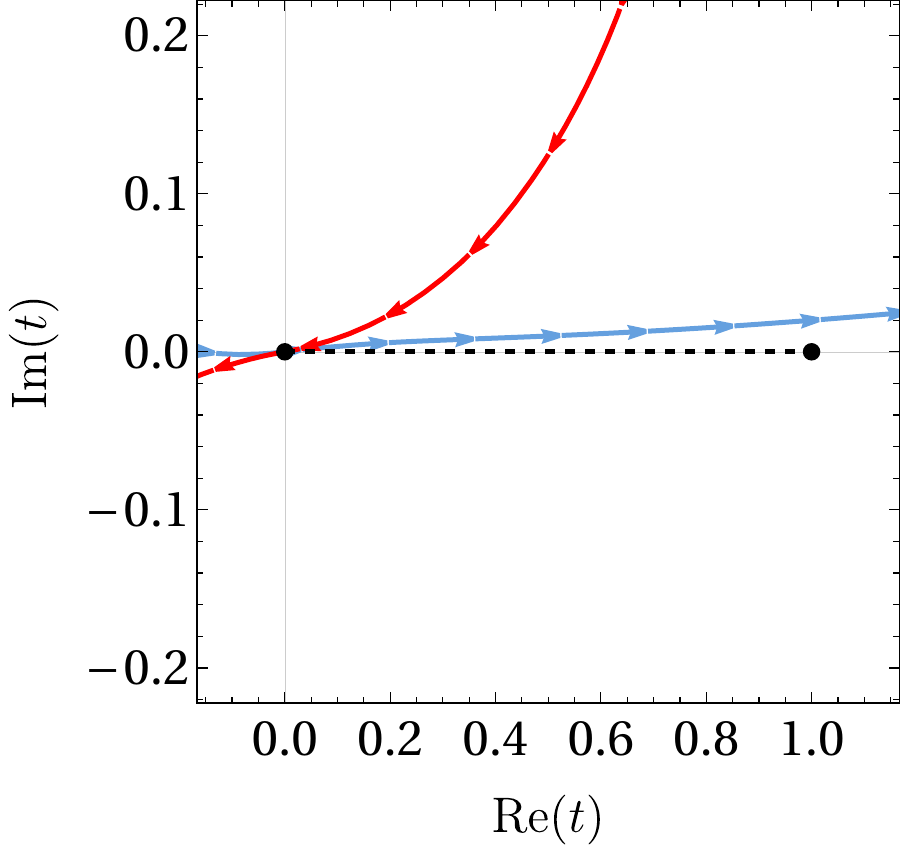}
		\hspace*{1cm}
		\includegraphics[width=0.4\textwidth]{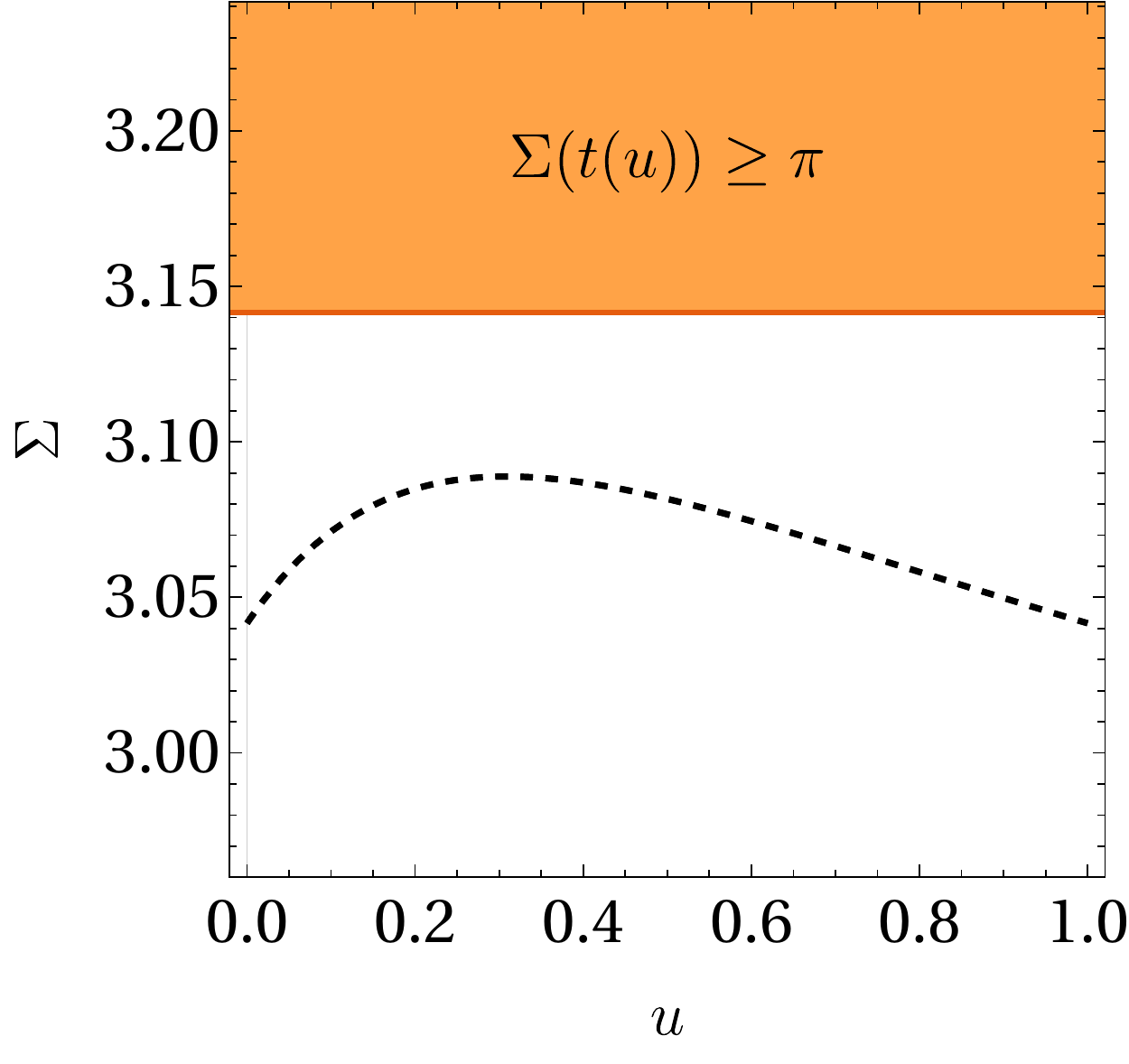}
		\caption{\textit{Left}: Complexified time plane for a solution with classical boundary conditions just off-shell of the expanding saddle point. Here $q_0=2$, $q_1=10$, and $\Lambda=3$, and the perturbation about the expanding saddle is taken to be $N=2-i/10$. Black dots reside at $t=0$ and $t=1$, indicating the end points of the time path specifying the metric. The blue directed line is the line of maximal angle, and the red directed line is that of minimal angle, so an allowable path must reside in between under forward time evolution -- recall we want $\Re(t)$ to grow monotonically from $0$ to $1$. How those minimal and maximal angle directed lines are found is explained in section \ref{subsec:minmaxcurves}. We can see that the point $t=1$ is reachable while staying below the blue line, for example using the black dashed path, which is simply on the real-$t$ axis here. \textit{Right}: We explicitly verify that the dashed path is indeed allowable, by plotting the full $\Sigma$ along it. It everywhere stays below the limiting value $\pi$, indicated by the orange line.}
		\label{fig:expansionex}
	\end{figure}

	To assess the allowability of (off-shell) metrics everywhere else in the complex lapse plane, in particular around the saddle points and along the steepest descent contours, we have to resort to the methods described in section \ref{subsec:methods}, which involve numerical computations.
	Specifically, we allow the time path from $t=0$ to $t=1$ to be deformed to any simply connected smooth path in the complexified time plane, while leaving the end points (where the boundary conditions are imposed) fixed. We would like to know if at least one path exists such that the resulting metric is allowable. One of the methods used (cf.~section \ref{subsec:minmaxcurves}) is to note that, at each point in the complex $t$ plane, there is a range of directions for which the LSKSW bound is satisfied, i.e., for which $\Sigma < \pi$. Starting from $t=0$, we can follow both the path where at each point we choose the maximum angle allowed and the path of minimum angle. Any allowable paths must reside in between these two lines. In figure~\ref{fig:expansionex}, paths must have an angle smaller than the blue curve and larger than the red curve, and we can see that the allowed region contains the point $t=1$, and hence one may reach it via an allowable path. This example corresponds to a lapse value just off-shell of the expanding saddle point. An example of an allowable path is given by the dashed line in the figure (just the real straight line in that example), and in the right panel we plot $\Sigma$ in order to explicitly verify that this geometry is indeed allowable.
	
\begin{figure}
		\centering
		\includegraphics[width=0.45\textwidth]{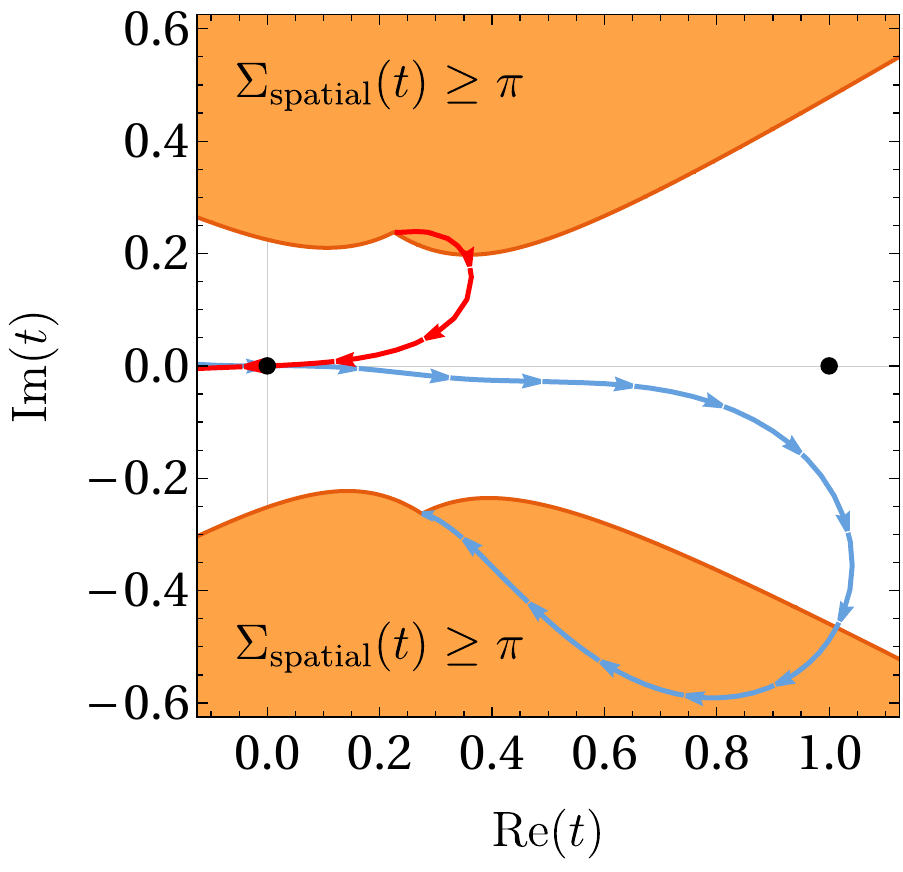}
		\caption{Same plot as on the left-hand side of figure~\ref{fig:expansionex}, except now the lapse is taken to have the value $N=4-i/10$, which is just off-shell of the bouncing saddle point. We added shaded regions, which correspond to non-allowable metrics according to the spatial LSKSW criterion alone, i.e., in these regions $3|\mathrm{Arg}[q(t)]|\geq \pi$. Time paths respecting the full LSKSW bound must reside in between the blue and red directed lines, hence it is impossible to reach $t=1$ in this example, and the metric is thus not allowable. In particular, a path reaching $t=1$ necessarily has to cross the blue line, where we know $\Sigma=\pi$.}
		\label{fig:bounceex}
	\end{figure}

	In contrast, for a value of the lapse just off-shell of the bouncing saddle point, we find a different result, as illustrated in figure~\ref{fig:bounceex}. There, we see that it is impossible to draw a simply connected time path from $t=0$ to $t=1$ while remaining below the blue curve and above the red one and while avoiding regions where the LSKSW bound is violated already from spatial part of the metric (which is independent of the angle of the path). Therefore, we may already expect the bouncing saddle point to be more difficult to approach along allowable metrics.
	
\begin{figure}
		\centering
		\includegraphics[width=0.45\textwidth]{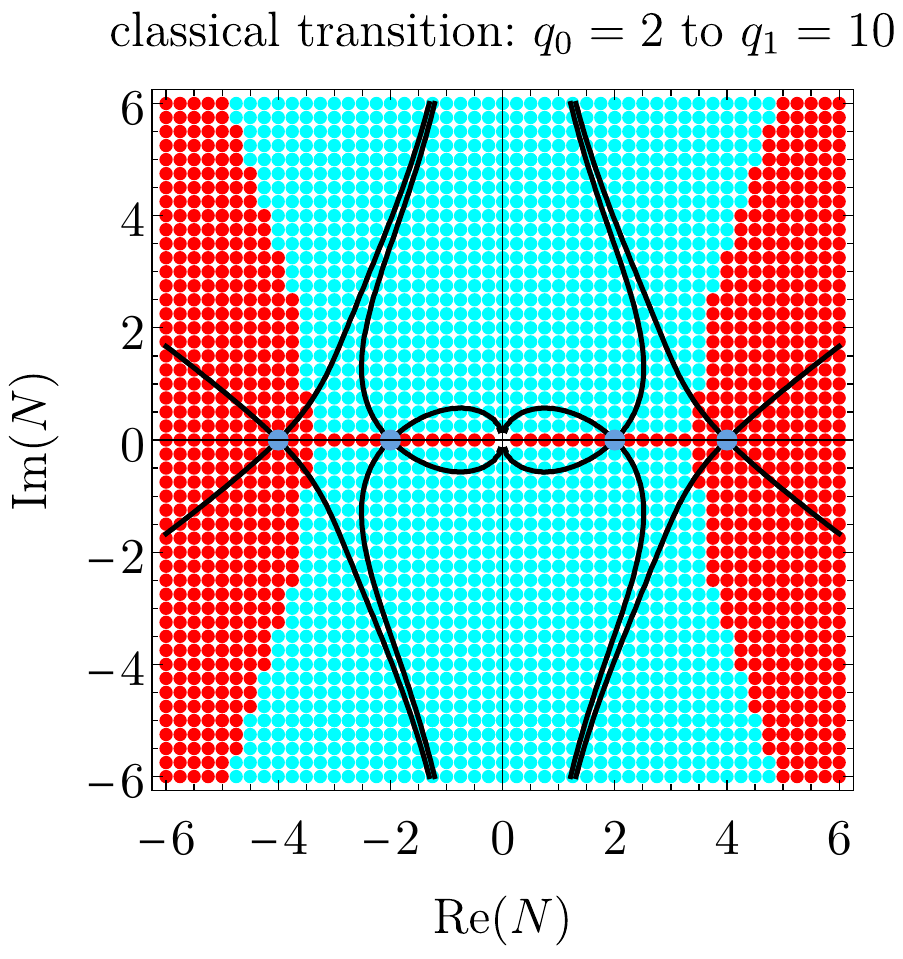}
		\caption{Complex lapse plane for the classical transition from a small scale factor, $q_0=2$, to a larger scale factor, $q_1=10$, with $\Lambda=3$. Saddle points (at $N_\mathrm{SP}=\pm 2,\pm 4$) are in blue, and contours of steepest ascent and descent are in black. The inner saddle points, corresponding to expanding solutions, are reachable via allowable complex metrics (in cyan), while the outer saddle points, corresponding to classical bouncing solutions, appear not to be reachable as they are surrounded by red -- non-allowable -- points. However, as argued in the core text, this is most likely an artifact of our metric ansatz, which proves to be too restrictive in the present case.}
		\label{fig:classical_transition}
	\end{figure}

	Using all the tools explained in section \ref{subsec:methods}, such as the lines of minimal and maximal angles for the time path (for which we gave a couple of examples above), we can determine the allowability of every point in the complexified lapse plane; see figure~\ref{fig:classical_transition} for an example of full results. We find that the inner (expanding) saddle points are always reachable by allowable complex metrics, i.e., they are surrounded by the domain of allowability, while the outer (bouncing) saddle points appear not to be reachable. We find that this is always true when the final scale factor is large enough, $q_1 \gtrsim 2$. This effect is, however, presumably due to the restrictive nature of our metric ansatz (motivation for this will be given shortly). This may not mean that bouncing saddles are not reachable via any off-shell contours, but rather that we are forced to depart from the minisuperspace steepest descent contour if we want to reach them. For example if we consider the following family of metrics for $\varepsilon\in(0,1)$,
	\begin{equation}
		\dd s^2=-(1-i\varepsilon)^2\frac{N^2}{\bar{q}(t)}\dd t^2+\bar{q}(t)\dd\Omega_{(3)}^2\,, \label{offcontour}
	\end{equation}
	and evaluate them on the outer saddle point,
	then we see that they form a continuous path in the space of off-shell metrics, which are all allowable\footnote{Indeed, whenever the lapse is real and taking the time path to be the real straight line, one finds $\Sigma(t)=\abs{\mathrm{Arg}[-(1-i\varepsilon)^2]}=\abs{\pi-\arctan[2\varepsilon/(1-\varepsilon^2)]}$, which is always strictly smaller than $\pi$ for any value of $\varepsilon\in(0,1)$.} and such that the limit $\varepsilon\to0^+$ of this continuous path is the bouncing saddle point metric. These metrics have a weighting that depends on $\varepsilon$, and one can then link them to an integration contour that asymptotes to the Euclidean axis at large lapse values, though such an integration contour will not be a steepest descent contour.
	
	Note in passing that the non-singular bouncing saddle point geometry does not violate the null energy condition (NEC) -- though it saturates it -- hence satisfying the NEC does not guarantee metric allowability.\footnote{The weak energy condition (WEC) is violated, however. Satisfying the WEC may well imply allowability, but the converse is not generally true \cite{Visser:2021ucg}.} In fact, the NEC is also saturated for the inner saddle points, and it is violated for the off-shell geometries that reside on the real-$N$ line in between inner and outer saddles. The whole real-$N$ line saturates the LSKSW bound, but a large interval of points in between the inner and outer saddles can be infinitely close to satisfying the LSKSW bound, yet violate the NEC. So once again, there appears to be no equivalence between metric non-allowability and NEC violation.
	
\begin{figure}
		\centering
		\includegraphics[width=0.45\textwidth]{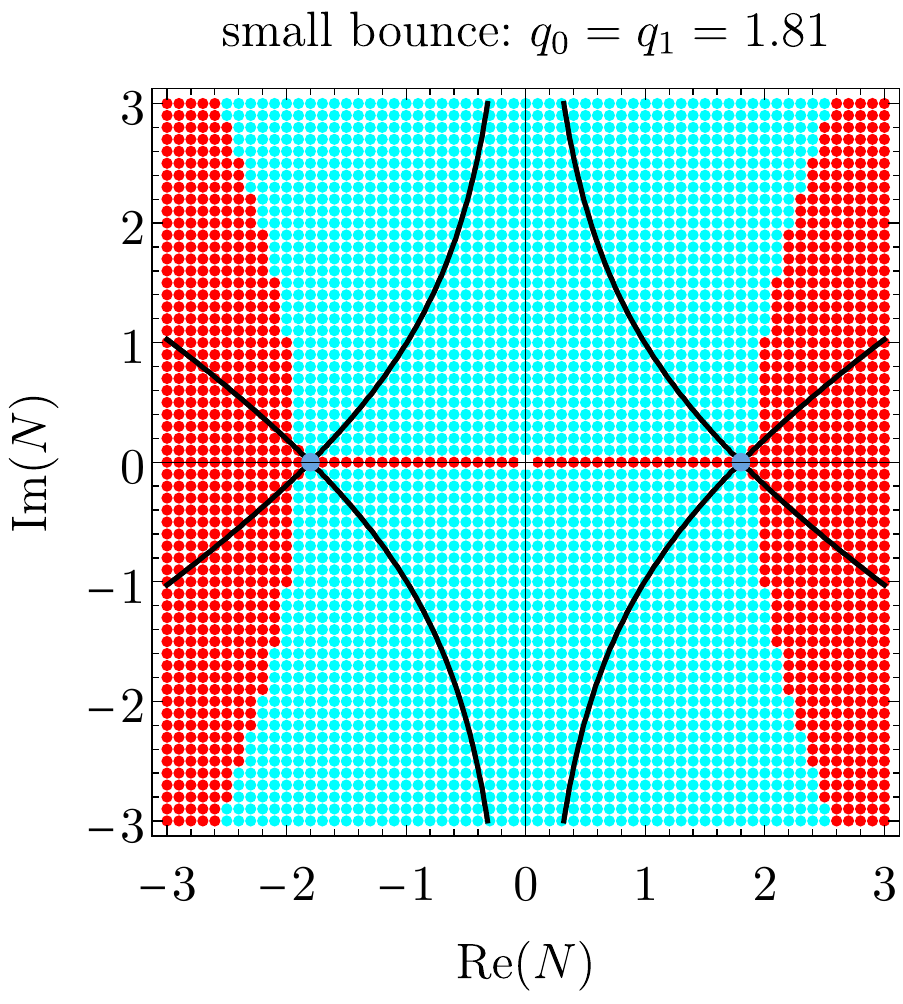}
		\caption{Classical transition with $q_0=q_1=1.81$. The saddle points (in blue) -- which correspond to classically bouncing backgrounds -- are reachable by allowable metrics along the steepest descent contours (in black).}
		\label{fig:smallbounce}
	\end{figure}

	If we now consider a bounce where the initial and final scale factor values are small, then we find that the classically bouncing saddle point solutions are in fact reachable through allowable minisuperspace metrics along the steepest descent contour. An example with $q_0=q_1=1.81$ (and saddle points at $N_\text{SP}=\pm1.8$) is given in figure~\ref{fig:smallbounce}. We find that this is a generic property when both $q_0$ and $q_1$ are smaller than $2$. This is surprising as it does not appear physically reasonable to have reachable `small bounces', yet non-reachable `large bounces', because the geometries are effectively the same -- their likeliness could be different, but this has nothing to do with the allowability criterion (and thus with their `reachability'). That is why we were led to conclude that the apparent `non-reachability' of the large bouncing saddle points is an artifact of the minisuperspace approximation.
	
	Let us elaborate on the apparent paradox with `large bounces' not being reachable. One could imagine a scenario where we have only two saddle points, which are time reversals of each other, with $q_0=1$ and $q_1\gg 1$. Then, we could glue the contracting saddle with the expanding one (both reachable according to the allowability criterion), but this would be exactly the same classical solution as a large bouncing saddle with $q_0=q_1\gg 1$ (which is apparently not reachable under the same criterion).
	There are two important distinctions, though: first, the transition amplitude for these scenarios is not the same since the scenario with a contraction and an expansion implies a measurement being made when the universe is indeed small, while there is no such prior knowledge for a `large bounce'. Second, if we imagine an off-shell approach to these saddle points with a slight imaginary part to the lapse -- all real saddles are on the boundary of the LSKSW bound anyway -- then the solutions are actually no more equal due to the imaginary part of $q$ (and thus their LSKSW functions differ). These distinctions resolve the paradox.
	
	These examples serve to stress that the minisuperspace model is often over-restrictive when considering the path integral approach to early universe models, and they also highlight the difficulty in finding appropriate contours of integration for the gravitational path integral. Even though the mathematically most natural procedure would be to look for steepest descent paths, we can see here quite explicitly that such paths may conflict with the criterion for allowability of complex metrics. In such cases, it is then important to find out if suitable integration contours, like the one indicated in \eqref{offcontour}, exist and if the saddle points in question are reachable. We will find further examples of this subtlety in section \ref{sec:no-boundary}.

	%%%%%%%%%%%%%%%%%%%%%%%%%%%%%%%%%%%%%%%%%%%%%%%%%%%%%%%%%%%%%%%%%%%%%%%%%%%%%%%%%%%%%%%%%%%%%%%%%%%%%%%
	
	\section{Quantum bounces}
	\label{sec:quantumBounces}
	
	Quantum theory allows for dynamical evolution in situations where the classical theory would forbid any transitions. For instance, a particle may tunnel across a potential barrier even if it does not possess enough kinetic energy to overcome the barrier. Such classically forbidden evolutions are typically unlikely, but possible. They may be usefully described by motions in complex time \cite{Bender:2008fr,Turok:2013dfa,Bramberger:2016yog} or, in the gravitational context, by complex metrics. A prominent example is provided by Coleman-De Luccia instantons, which describe the nucleation of a bubble of true vacuum in a false vacuum region of the universe \cite{Coleman:1980aw}. These transitions are clearly allowable, as the corresponding metrics are purely Euclidean. But one may also envisage even more momentous transitions, such as a universe (or perhaps just a region thereof) tunneling from a contracting phase to an expanding phase, thus circumventing the big bang by moving around it in complexified field space. Examples of such transitions have been presented in \cite{Bramberger:2017cgf,Boyle:2018tzc,Boyle:2021jej} and they involve genuinely complex metrics. Here we would like to assess whether such transitions are allowable or not. 
	
	The transitions presented by Bramberger et al.~\cite{Bramberger:2017cgf} all involved a complexified scalar field in addition to a complex metric, and as such they would immediately be ruled out by the assumptions going into the LSKSW criterion. However, one may easily envisage analogous situations where the scalar field remains real (for some transitions in \cite{Bramberger:2017cgf} the scalar is almost real), or where it is absent altogether. We refer to the paper \cite{Bramberger:2017cgf} for explicit numerical examples of quantum bounce solutions. For our purposes here, it is helpful to realise that the prototype for the transitions in \cite{Bramberger:2017cgf} is a kind of back-to-back gluing of no-boundary geometries, which we can analyse straightforwardly.
	
\begin{figure}[t]
		\centering
		\includegraphics[width=0.45\textwidth]{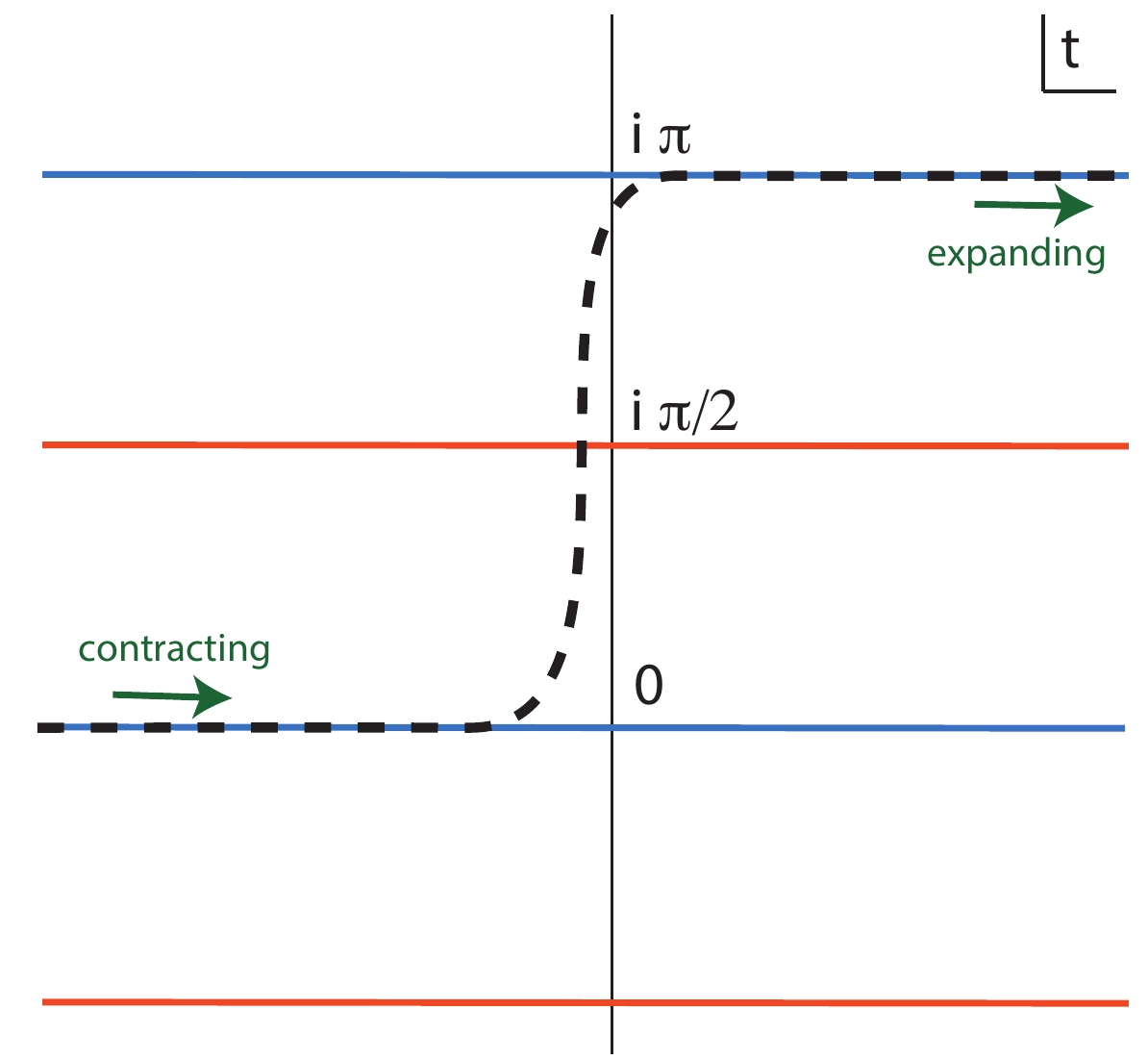}
		\caption{Lines of real (in blue) and pure imaginary (in orange) scale factor values in the complexified time plane, for $a(t)=\cosh(t)$. In this figure, real Lorentzian time is horizontal and Euclidean time vertical. One can model a quantum bounce (dashed black line) by linking a contracting universe to an expanding one at a different, parallel line of real field values, but such a path necessarily passes through a line on which the scale factor is imaginary (orange line).}
		\label{Fig:quantumbounces}
	\end{figure}

	If the vacuum energy $\Lambda=(D-1)(D-2)H^2/2$ is constant, either because it consists of a cosmological constant or of a scalar field at an extremum of the potential, the scale factor (in a spatially closed universe) is given by
	\begin{equation}
		a(t) = \frac{1}{H} \cosh(Ht)\,. \label{metricqb}
	\end{equation}
	This solution already bounces at the classical level, hence if this were the true solution, we would have no need for a quantum bounce. However, a classical bounce requires extremely fine tuned initial conditions and cannot be regarded as realistic. It is far more likely that the initial conditions depart somewhat from the ones leading to the solution \eqref{metricqb}, in particular if anisotropies are present. Then in the contracting phase the true classical solution deviates more and more from the pure $\cosh(Ht)$ solution, ending in a big crunch (as anisotropies grow much faster than the homogeneous curvature term that is required to induce a classical bounce; see \cite{Bramberger:2019zez} for an analysis of when a bounce can occur). We can then circumvent this crunch by choosing a path in the complexified time plane that links up with an expanding solution (see figure~\ref{Fig:quantumbounces}), avoiding the crunch and providing a possible channel for the contracting universe to reach an expanding phase.
	
	More precisely, in the contracting part of the solution the imaginary part of $t$ vanishes and $t$ is negative. After the transition one wishes to follow a line with $\Im(Ht)=\pi$ (or in general one with $\Im(Ht)=n\pi$, $n \in \mathbb{Z}$), along which we can find the expanding branch for positive $t$ (for odd $n$ the scale factor is negative, but this is immaterial as the metric contains $a^2$). The coordinate value $Ht=i\pi/2$ would then be the no-boundary point, where the manifold degenerates to a point. We can regularise this by transitioning slightly away from $\Re(t)=0$, i.e., we can interpolate between contracting and expanding branches along a line $Ht=\pm\epsilon+i\upsilon$ with $\upsilon$ spanning $[0,\pi]$ (here $\epsilon$ is a small constant positive real number). This line, or indeed any other path connecting the two branches, must pass through the line $Ht=x+i\pi/2$, $x\in(-\infty,\infty)$, but along that line $a=H^{-1}\cosh(x+i\pi/2)=iH^{-1}\sinh(x)$. Thus, the scale factor squared $a^2=-H^{-2}\sinh^2(x)$ is a negative real number along that line, and hence on that line we have, following from $\Sigma(t)\geq\Sigma_\textrm{spatial}(t)=(D-1)|\mathrm{Arg}[a(t)^2]|$ for a $D$-dimensional FLRW spacetime,
	\begin{equation}
		\Sigma \geq (D-1) \pi\,,
	\end{equation} 
	so that such solutions are not allowable. In more general situations, the lines of real field values are somewhat deformed, but if the dynamics contains an attractor (such as inflation), then lines of real field values are asymptotically parallel to real-time axis. In attempting to interpolate between two such lines one will then once again pass a locus where the scale factor is pure imaginary [in fact it is already enough if its argument is larger than $\pi/(D-1)$], and then the metric is seen not to be allowable. 
	
	As an aside, we can now see that the no-boundary solutions with multiple spheres attached, discussed in section \ref{subsec:noboundarynspheres} as motivation, are not allowable for the same reason: each time a new sphere gets attached, the regularised time path that passes around the zeroes of the scale factor (i.e., following the dotted lines in figure~\ref{fig:no-boundaryManySpheres}) crosses a line where the scale factor is purely imaginary and where the LSKSW bound is violated.
	
	Let us now comment in more detail on why (with hindsight) it is a good thing that the quantum bounces are excluded from the path integral. The prototype example described above closely resembles the gluing together of two no-boundary geometries. Studies of these geometries (e.g., \cite{Bousso:1996au,Feldbrugge:2017mbc}) have shown that they come in two types, related by complex conjugation (i.e., corresponding to whether one links the contracting solution with an expanding one at either positive or negative imaginary time values). If the solution exhibits a negative weighting, then it occurs with exponentially suppressed probability, of order $e^{-|\mathcal{I}(S^4)|/\hbar}$, where $\mathcal{I}(S^4)$ is the action of the 4-sphere with radius $1/H$. However, in that case fluctuations around the solution obtain positive weighting proportional to the size of the fluctuations squared. In other words, fluctuations obtain an inverse Gau\ss{}ian probability distribution, ultimately destroying the consistency of the model. The complex conjugate solution has stable perturbations, obeying a Gau\ss{}ian probability distribution, but in that case the weighting is positive, now of order $e^{+|\mathcal{I}(S^4)|/\hbar}$, implying that such solutions would occur with higher probability than classical evolution. Quantum transitions of this sort would thus be the norm, and not the exception, in clear contradiction with our actual experience. We may summarise by saying that quantum bounces built out of approximate complexified de Sitter solutions are thus unstable, either at the background or at the perturbative level, and their exclusion gives support to the validity of the LSKSW criterion. 
	
	A different type of quantum bounce has been explored by Boyle, Turok, and collaborators \cite{Boyle:2018tzc,Boyle:2021jej} and goes by the name of two-sheeted universe (or CPT-symmetric universe). In this model, no inflationary phase is assumed, and in fact in the approach of the big bang the universe is dominated by radiation all the way. The metric then approaches the form
	\begin{equation}
		\dd s^2 = \eta^2 (-\dd\eta^2 + \dd x^2 + \dd y^2 + \dd z^2)\,,\label{eq:metricradconf}
	\end{equation}
	where $\eta$ denotes the conformal time, and it was assumed that the spatial sections of the universe are flat. If $\eta$ remains real valued, this geometry describes two cones, one for $\eta>0$ and one for $\eta<0$, joined at their tips where $\eta$ vanishes. The idea of \cite{Boyle:2018tzc} is to pass around the big bang at $\eta=0$ by taking a path in the complexified time plane. Suppose this path is just a semi-circle at a small radius, i.e., $\eta=r e^{i\theta}$ with $\theta$ spanning $[0,\pi]$ and $r>0$ kept constant. Then at $\theta=\pi/2$ we have $\eta=ir\Rightarrow\eta^2=-r$ and consequently, from the spatial part of \eqref{eq:metricradconf} and the LSKSW bound, we find $\Sigma \geq (D-1) \pi$,
	which violates the LSKSW bound in all dimensions greater than $2$. More generally, the locus of negative values of $\eta^2$ emanates radially from $\eta=0$ and is necessarily crossed as we pass from $\eta>0$ to $\eta<0$ by any continuous path avoiding the singularity at $\eta=0$. Hence it would seem that the two-sheeted universe is not allowable. We should point out a possible caveat: in the works on the two-sheeted universe, it is emphasised that in the approach to the big bang conformal symmetry might become manifest \cite{Gielen:2015uaa,Gielen:2016fdb,Boyle:2021jaz}. If this symmetry, or any other, becomes exact then this would modify some of the assumptions that go into the derivation of the LSKSW bound, as the matter contributions would also have to obey the symmetry in question. Since this would limit the possible matter types that have to possess convergent actions, this would allow for more general metrics than those satisfying the LSKSW criterion. In particular, for exact conformal symmetry the scale factor of the universe is not restricted. It will be of interest to pursue this question in future work.

	%%%%%%%%%%%%%%%%%%%%%%%%%%%%%%%%%%%%%%%%%%%%%%%%%%%%%%%%%%%%%%%%%%%%%%%%%%%%%%%%%%%%%%%%%%%%%%%%%%%%%%%
	
	\section{No-boundary wave function}
	\label{sec:no-boundary}
	
	Our universe underwent several phases of evolution, in which different matter components dominated the dynamics, e.g., the phases of radiation, matter or dark energy domination. In order to describe the cosmic evolution during a given phase, we must specify initial conditions, which in fact are the final conditions of the prior phase. Hawking argued \cite{Hawking:1981gb} that the implied infinite regression can be avoided if the initial phase of the universe was described by a compact, regular geometry, so that there would be no need to specify further conditions at the beginning. In fact such a no-boundary geometry would effectively replace the big bang singularity and provide initial conditions for the universe. 
	As implied by the singularity theorems of Penrose and Hawking (e.g., \cite{Hawking:1970zqf}), a Lorentzian geometry is necessarily singular at the big bang and cannot satisfy the advocated compactness and regularity requirements. However, a Euclidean (or more generally complex) geometry can. This led Hartle \& Hawking to propose that the wave function of the universe should be given by a (formally Euclidean) path integral summing over compact and regular metrics \cite{Hartle:1983ai}. Precisely defining such a path integral has been a long standing goal, due to various complications, of which we will highlight two. The first is the conformal mode problem, namely the fact that the kinetic term for the scale factor of the universe has the opposite sign to the matter kinetic terms, obstructing the convergence of the Euclidean path integral. The second issue is that compactness is a condition on the scale factor, while regularity is a condition on the expansion rate -- since these are canonically conjugate variables, the uncertainty principle forbids one to impose both conditions simultaneously in the definition of the path integral. 
	
	In recent years, there has been significant progress on these issues, when the path integral is restricted to minisuperspace metrics, i.e., to non-generic metrics with (physically motivated) symmetries. For instance, the integrals can be defined on appropriate integration contours using Picard-Lefschetz theory \cite{Feldbrugge:2017kzv} (see also \cite{Halliwell:1988ik}). Also, it was found that imposing compactness leads to problems with the stability of perturbations \cite{Feldbrugge:2017fcc} (and consequently with the consistency of the calculations \cite{Feldbrugge:2017mbc}; for contrasting views see \cite{DiazDorronsoro:2017hti,PhysRevLett.121.081302,Feldbrugge:2018gin}), while imposing regularity in fact leads to a satisfactory definition of the no-boundary wave function \cite{Louko:1988bk,DiTucci:2019dji,DiTucci:2019bui,Lehners:2021jmv}.\footnote{Moreover, imposing a regularity/momentum condition is directly related to the gravitational path integral in asymptotically anti-de Sitter spacetimes via analytic continuation \cite{DiTucci:2020weq,Lehners:2021jmv}.} We will review both constructions below, from the point of view of the question of allowability of metrics. For the sum over regular metrics, we can prove that no-boundary geometries are allowable in arbitrary spacetime dimensions, that they always reside on the boundary of the domain of allowable metrics, and moreover that appropriate integration contours for the path integral exist. For the sum over compact metrics, the saddle points are allowable but not on the boundary of the allowable domain. Here, as we will explain, the restrictions implied by the criterion of allowability may in fact permit a definition of the path integral avoiding the above-mentioned issues with perturbative instabilities.
	
	\subsection{Sum over regular metrics}
	
	In the present section, we return to the minisuperspace model already used in section \ref{sec:classicalTransitions}, but now in arbitrary spacetime dimension $D$; in particular, the action is taken to be
	\begin{equation}
		S = \frac{1}{2}\int \dd^Dx\,\sqrt{-g}\left(R-2\Lambda \right)\,.\label{eq:actionD}
	\end{equation}
	Hence the spatial sections of the metric \eqref{metric} are taken to be $(D-1)$-spheres, i.e., with metric $\dd\Omega_{(D-1)}^2$. First, we should determine what condition we must put on metrics in order for them to be regular. For this purpose it is useful to look at the constraint/Friedmann equation (which follows from the variation of the action \eqref{eq:actionD} with respect to the lapse)
	\begin{equation}
		\frac{\dot{q}^2}{4N^2} + 1 = \frac{2\Lambda}{(D-1)(D-2)} q \,.
	\end{equation}
	Regularity of this equation at $q=0$ requires $\dot{q} = \pm 2Ni$. The presence of the imaginary unit $i$ implies that the expansion rate must be imaginary, or in other words, the metric must start out Euclidean near $q=0$. The choice of sign for $\dot{q}$ corresponds to the effective choice of Wick rotation. In order for perturbations (e.g., tensor perturbations) to be stable \cite{Feldbrugge:2017fcc}, we must choose the positive sign
	\begin{align}
		\left.\dot{q}\right|_\textrm{origin} = 2Ni \qquad \textrm{(regularity condition)}\,.
	\end{align}
	As discussed in section \ref{sec:classicalTransitions}, the path integral is sliced such that the time coordinate $t$ runs from $t=0$ to $t=1$. On the initial hypersurface $t=0$ we will thus impose the regularity (Neumann) condition above, and on the final hypersurface we will stick to a Dirichlet condition $q(t=1)=q_1$:
	\begin{align}
		\Psi(q_1) = \int {\mathcal D}N\int\limits_{\dot{q}=2Ni}^{q=q_1}{\mathcal D}q~\exp(\frac{i}{\hbar}S)\,.
	\end{align}
	In $4$ dimensions, this path integral has been studied in detail in \cite{DiTucci:2019bui,DiTucci:2020weq,Lehners:2021jmv}. There it was shown that the integral over $q$ can be performed by writing $q(t)=\bar{q}(t)+Q(t)$, where $\bar{q}$ denotes a solution to the equation of motion that follows from the variation with respect to $q$,
	\begin{equation}
		\ddot{q} = \frac{4\Lambda N^2}{(D-1)(D-2)} \,, \label{qeom}
	\end{equation}
	while $Q(t)$ is an arbitrary fluctuation satisfying the boundary conditions $\dot{Q}(t=0)=Q(t=1)=0$. In $4$ dimensions, the integral over $Q$ is a Gau\ss{}ian and simply yields a numerical factor. In general dimensions, this integral cannot be performed so easily, and we leave a full analysis for future work. We can nevertheless make progress by looking at solutions to the equation of motion, which for mixed Neumann-Dirichlet (ND) boundary conditions are explicitly given by
	\begin{equation}
		\bar{q}_\textrm{ND}(t)=\frac{2\Lambda N^2}{(D-1)(D-2)} (t^2-1) + 2Ni(t-1) + q_1\,.
	\end{equation}
	The path integral will then sum over a subset of these backgrounds, depending on the integration contour for $N$. We want to know whether the complexified lapse space has a special structure and which regions correspond to (non-)allowable metrics.
	
\begin{figure}[t]
		\centering
		\includegraphics[width=0.4\textwidth]{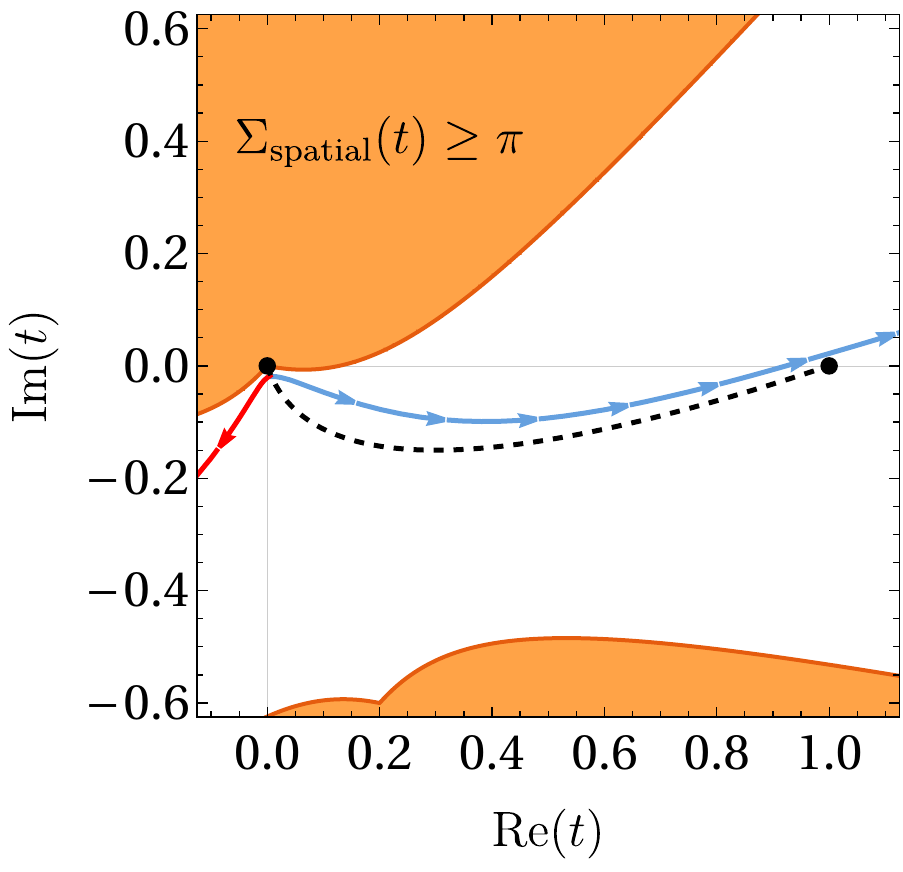}
		\hspace*{1cm}
		\includegraphics[width=0.4\textwidth]{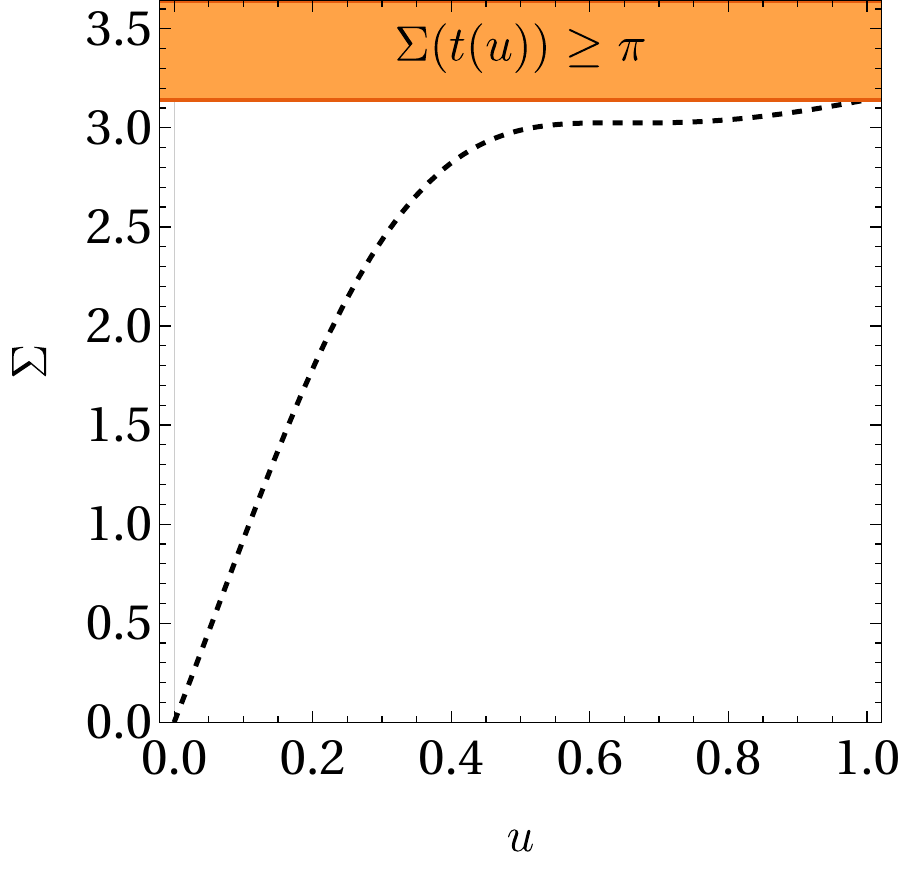}
		\caption{\textit{Left}: Complexified $t$ plane for the saddle point solutions with Neumann-Dirichlet boundary conditions. Here $q_1=10$ and $\Lambda=(D-1)(D-2)/2$. The shaded region corresponds to non-allowed metrics according to the spatial LSKSW criterion alone, i.e., in these regions $(D-1)|\mathrm{Arg}(q)|\geq \pi$. Black dots reside at $t=0$ and $t=1$, indicating the end points of the time path specifying the metric. The blue line is the line of maximal angle and the red that of minimal angle -- an allowable path must reside in between. How those lines are found is explained in section \ref{subsec:minmaxcurves}. We can see that the point $t=1$ is reachable while staying below the blue and above the red lines, for example using the black dashed path. \textit{Right}: We explicitly verify that the dashed path is indeed allowable, by plotting the full $\Sigma$ along it. It everywhere stays below the limiting value $\pi$, indicated by the orange line.}
		\label{fig:NDsaddle}
	\end{figure}

	The location of the no-boundary saddle points $N_\pm$ can be determined by asking that $\bar{q}_\textrm{ND}(t=0)|_{N_\pm}=0$, yielding
	\begin{equation}
		N_\pm = \frac{(D-1)(D-2)}{2\Lambda}\left(-i \pm \sqrt{\frac{2\Lambda}{(D-1)(D-2)} q_1 - 1} \right)\,.
	\end{equation}
	We should first verify that the saddle points are allowable. We can do this using the methods described in section \ref{sec:allowableMetrics}.
	Specifically, we allow the time path from $t=0$ to $t=1$ to be deformed. Starting from $t=0$, we follow both the path where at each point we choose the maximum angle allowed and the path of minimum angle, see \eqref{eq:argtpineq}. In figure~\ref{fig:NDsaddle} we can see that the allowed region contains the point $t=1$, and hence one may reach it via an allowable path. An example of such a path is given by the dashed line in the figure, and in the right panel we plot $\Sigma$ in order to explicitly verify that this geometry is indeed allowable. Explicitly, the path that we chose is given by 
	\begin{align}
		t(u)=e^{i\left(\alpha-\pi/2\right)} \left[u(2-u)\sin(\alpha)+iu^2\cos(\alpha)\right]\,, \quad \alpha \equiv - \mathrm{Arg}(N_+)\,, \quad 0 \leq u \leq 1\,.
	\end{align}
	Note that the calculation works in any dimensions, since one can choose a length unit such that $\Lambda = (D-1)(D-2)/2$, which reduces the calculations in all dimensions to a single calculation.
	
\begin{figure}[t]
		\centering
		\includegraphics[width=0.4\textwidth]{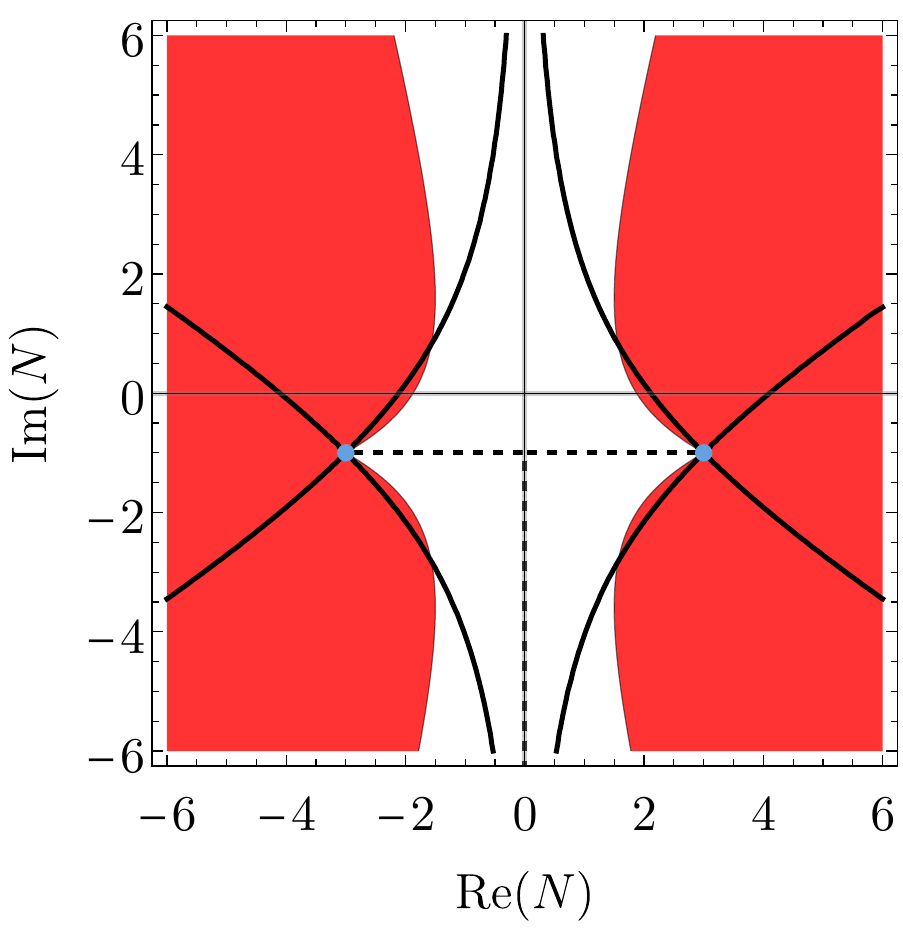}
		\hspace*{1cm}
		\includegraphics[width=0.4\textwidth]{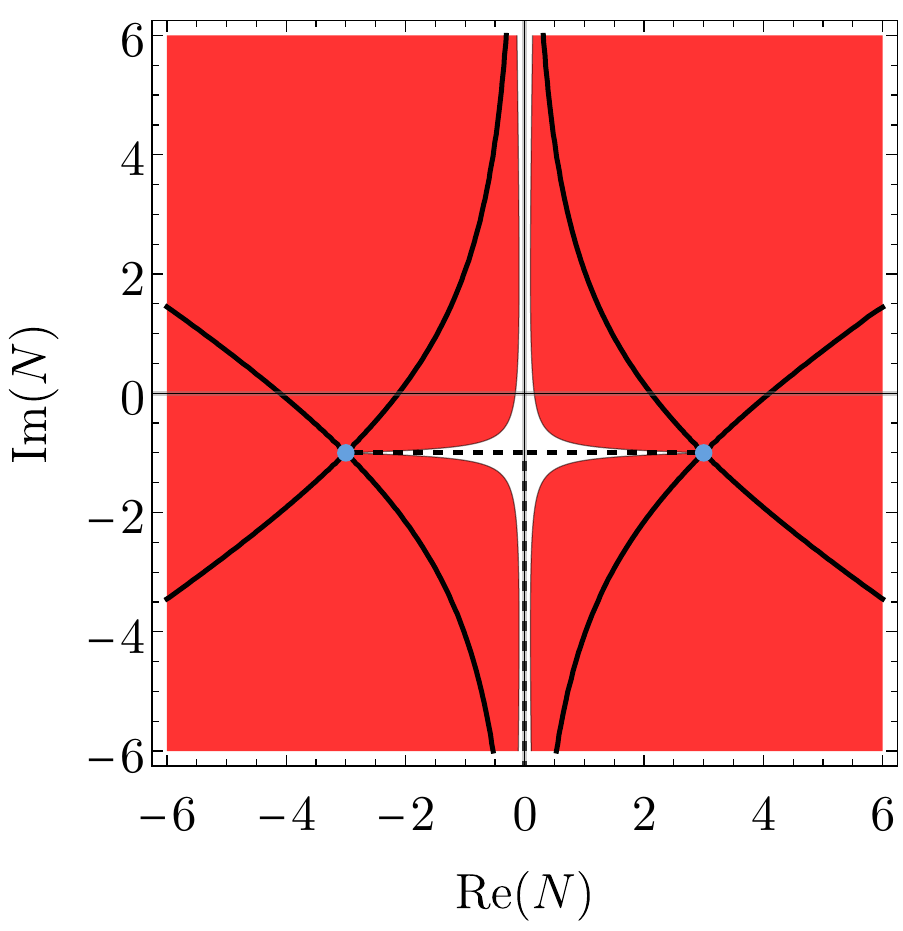}
		\caption{The red regions show the domain in the complexified lapse plane that is non-allowable according to the spatial LSKSW bound only, in $7$ dimensions (left panel) and in $100$ dimensions (right panel). Here we have chosen $\Lambda=(D-1)(D-2)/2$ and  $q_1=10$. In higher dimensions, the non-allowable region grows, leaving only a narrow cross-shaped region available, but the saddles (blue dots) remain at the boundary of the allowable domain. The full black curves depict the thimbles, and the dashed ones represent an example of T-shaped allowable integration contour.}
		\label{fig:7D}
	\end{figure}

	Having noticed that the saddle points are allowable, we may also establish that they are on the boundary of the allowable domain, as they are surrounded in some directions by non-allowable metrics. This was shown previously in \cite{Lehners:2021mah}, and we will generalise the arguments used there. In fact it is enough to consider only the spatial part of the metric. If this part already violates the allowability bound, then no change of time path can render the geometry allowable. As one may easily verify, the strongest constraint arises at $t=0$. Then let us look at values of the lapse near the saddle points, $N=N_\pm+\zeta$, for some $\zeta\in\mathbb{C}$ with $|\zeta|\ll 1$. Then $q(0) \approx \mp\,2 \zeta \sqrt{\frac{2\Lambda}{(D-1)(D-2)}q_1-1}$. We are interested in the case where $q_1 > \frac{(D-1)(D-2)}{2\Lambda}$, corresponding to a universe larger than the Hubble radius. Then the corresponding metric has a chance of being allowable only if $(D-1)|\mathrm{Arg}(q)| <\pi$, which leaves the wedges
	\begin{equation}
		-\frac{\pi}{D-1}< \mathrm{Arg}(\zeta) < \frac{\pi}{D-1} \,\,\, (\textrm{for }N_-)\quad \textrm{and} \quad  \frac{(D-2)\pi}{D-1}< \mathrm{Arg}(\zeta) < \frac{D \pi}{D-1}\,\,\, (\textrm{for }N_+)\,.
	\end{equation}
	Hence we can see that starting from the saddle points only a narrow wedge (around the direction in which the other saddle resides) is potentially allowable. Numerically, we can plot the disallowed regions in the entire complexified lapse plane, see the examples in figure~\ref{fig:7D}. For these plots we again use only the spatial part of the metric, near $t=0$, which implies that we can exclude regions where
	\begin{equation}
		(D-1)\abs{\mathrm{Arg}\left(q_1-2Ni-\frac{2\Lambda}{(D-1)(D-2)}N^2\right)}  \geq \pi\,.
	\end{equation}
	As the figure shows, in high dimensions this only leaves a very narrow cross-shaped region including the Euclidean axis and the thickened line segment joining the two saddle points. Moreover, the thimbles are now to a significant extent in the non-allowable region. However, in the following we will show that T-shaped integration contours over allowable metrics (running up the imaginary axis and then extending sideways to the two saddle points; shown by the dashed, black lines in figure~\ref{fig:7D}) remain. Hence the path integral may in fact be defined as a sum over allowable metrics.
	
\begin{figure}[t]
		\centering
		\includegraphics[width=0.45\textwidth]{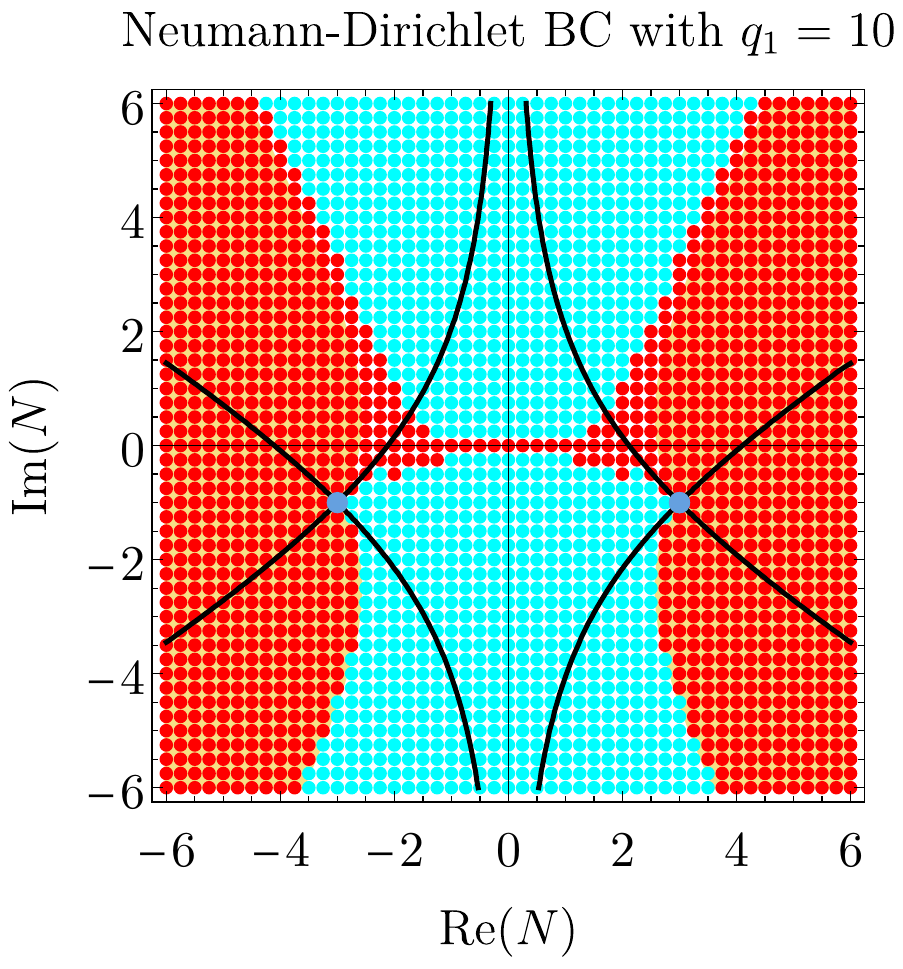}
		\caption{A numerical plot of the allowed (cyan) and disallowed (red) metrics in the complexified lapse plane satisfying initial Neumann and final Dirichlet conditions, as appropriate for a momentum space definition of the no-boundary wave function. Here $D=4$, $\Lambda=3$, $\dot{q}|_{t=0}=+2Ni$, and $q(t=1)=10$. The shaded orange region is the disallowed region discussed and plotted in figure~\ref{fig:7D}, though in 4 dimensions here. The saddle points (in blue) reside at the boundary of the allowable domain. The steepest ascent/descent lines (in black) emanating from the saddle points are seen to significantly impinge on the non-allowable region. Nevertheless, integration contours residing entirely in the lower half plane remain allowed.}
		\label{fig:ND4D}
	\end{figure}

	Before doing this, let us briefly specialise to $4$ dimensions, where we can use both our optimisation techniques and the method of maximal-minimal angles to determine numerically which regions of the complex lapse plane are actually allowable; see figure~\ref{fig:ND4D} for the results. This figure shows that the spatial bound used above already captures the non-allowable domain pretty well, but that in actuality it is somewhat larger, implying that for some geometries for which the spatial part of the metric would be allowable, no suitable time path can be found along which $\Sigma$ remains below $\pi$ everywhere. In particular, the region on and surrounding the real lapse axis is revealed not to be allowed. Importantly, we see though that the metrics along the negative Euclidean axis and those along the line segment between the two saddles remain allowable. We will now analytically show that this property remains true in all dimensions $D$.
	
	Thus, going back again to arbitrary dimension $D$, we will demonstrate that within the mentioned T-shaped region appropriate time paths exist that render the metrics allowable. Here we choose units such that $\Lambda=(D-1)(D-2)/2$. We will start with metrics on the imaginary axis, say $N=iN_\mathrm{E}$. Then the metric takes the form
	\begin{equation}
		\left.\bar{q}_\textrm{ND}(t)\right|_{N=iN_\mathrm{E}} = q_1 + N_\mathrm{E}^2 (1-t^2) + 2N_\mathrm{E}(1-t)\,.
	\end{equation}
	Here we can simply stick to the real time path where $t$ spans $[0,1]$. Then, given that the metric is real, all we need to verify is that it remains positive at all times. The only term that can give a negative contribution is the last term, when $N_\mathrm{E}<0$. The most stringent bound then arises when $t=0$, but then we find that the minimum of $q$ occurs when $N_\mathrm{E}=-1$. This minimal value is $q_\textrm{min}=q_1-1$, and it is positive as long as $q_1>1$, i.e., as long as the final hypersurface is larger than the waist of the de Sitter hyperboloid. This is in any case needed in order to obtain a classical universe.
	
	For lapse values on the segment in between the two saddles, we have to choose a different time path $t(u)\in\mathbb{C}$, as $q$ becomes complex along the real-$t$ line and thus in high-enough dimensions the LSKSW bound would surely be violated. However there exists a real-$q$ curve that connects $t=0$ to $t=1$. If we follow that curve, then the LSKSW bound will always be satisfied or at most saturated, because on this path the spatial part of the bound is zero and then we can have a contribution of at most $\pi$ from the path-dependent part $|\mathrm{Arg}[-N^2t'(u)^2/q(t(u))]|$. More explicitly, on the line $N=\mathcal{N}-i$, $\mathcal{N}\in(-\infty,\infty)$, and with $t(u)=v(u)+iw(u)$, $v(u),w(u)\in\mathbb{R}$, we have
	\begin{align}
		\left.\bar{q}_\textrm{ND}(t(u))\right|_{N=-i+\mathcal{N}}=&~q_1-\left(v(u)-1\right)^2\nonumber\\
		&+\big[2\,\mathcal{N}\left(2v(u)-1\right)+w(u)\big]w(u)+\mathcal{N}^2\left(v(u)^2-w(u)^2-1\right)\nonumber\\
		&+2i\big[\mathcal{N}v(u)+w(u)\big]\big[1-v(u)+\mathcal{N}w(u)\big]\,.
	\end{align}
	The imaginary part vanishes on the lines $w=-\mathcal{N}v$ and $w=(v-1)/\mathcal{N}$ (as long as $\mathcal{N}\neq 0$). These lines respectively run through $t=0$ ($v=w=0$) and $t=1$ ($v=1$ and $w=0$), and moreover they cross at $v=1/(\mathcal{N}^2+1)$, $w=-\mathcal{N}/(\mathcal{N}^2+1)$. Noting that $q(t=0)=q_1-1-\mathcal{N}^2$ must remain positive for the LSKSW bound not to be violated requires $|\mathcal{N}|<\sqrt{q_1-1}$, which is precisely the condition that the lapse must lie in between the two saddles. With this condition (and if $\mathcal{N}\neq 0$), it is straightforward to show that on the line segments $w=-\mathcal{N}v$ for $0\leq v\leq 1/(\mathcal{N}^2+1)$ and $w=(v-1)/\mathcal{N}$ for $1/(\mathcal{N}^2+1)\leq v\leq 1$, $\bar{q}_\mathrm{ND}$ is not only real, but also positive (we are restricted to $0 \leq v \leq 1$ so that $\Re(t(u))\in[0,1]$). For completeness let us check the value of the LSKSW bound. On the first line segment, we can take $t(u)=u-i\mathcal{N}u$, so that $t'(u)=1-i\mathcal{N}$ and hence $\mathrm{Arg}[-N^2 t'(u)^2/\bar{q}_\mathrm{ND}(t(u))]=\mathrm{Arg}[(1+\mathcal{N}^2)^2]=0$. On the second line segment, we can take $t(u)=u+i(u-1)/\mathcal{N}$, so that $t'(u)=1+i/\mathcal{N}$ and hence $\mathrm{Arg}[-N^2t'(u)^2/\bar{q}_\mathrm{ND}(t(u))]=\mathrm{Arg}[-(\mathcal{N}+1/\mathcal{N})^2]=\pi$. Thus on the second segment the LSKSW bound is saturated. By slightly smoothing out the path, just as at the saddle point itself, one can ensure that the LSKSW bound is satisfied. Obviously, the higher the dimension, the closer one has to stay to the real $q$ segment described above.
	
	Therefore, we can understand why the T-shaped region remains allowable up to arbitrarily large dimensions, and the path integral may be defined with two lapse integration contours that remain inside the allowed region from $-i\infty$ up and then left or right respectively towards the two saddle points (e.g., the T-shaped contours shown in figure~\ref{fig:7D} by the dashed black lines). At the saddle points, the integration contours come to an end, as they then run into a disallowed region. Presumably, this changes the semi-classical approximation to the wave function insignificantly, though it would of course be of interest to investigate this question further.

	\subsection{Sum over compact metrics}
	
	In the previous sub-section, we saw that the no-boundary proposal, when defined in momentum space, is compatible with a restriction to allowable metrics only, in arbitrary spacetime dimension. This is significant, as the no-boundary proposal relies on the use of complex metrics. In the present section, we would like to perform a similar analysis for the no-boundary wave function defined in field space, i.e., as a sum over compact metrics. We will work exclusively in $4$ dimensions. 
	
\begin{figure}[t]
		\centering
		\includegraphics[width=0.45\textwidth]{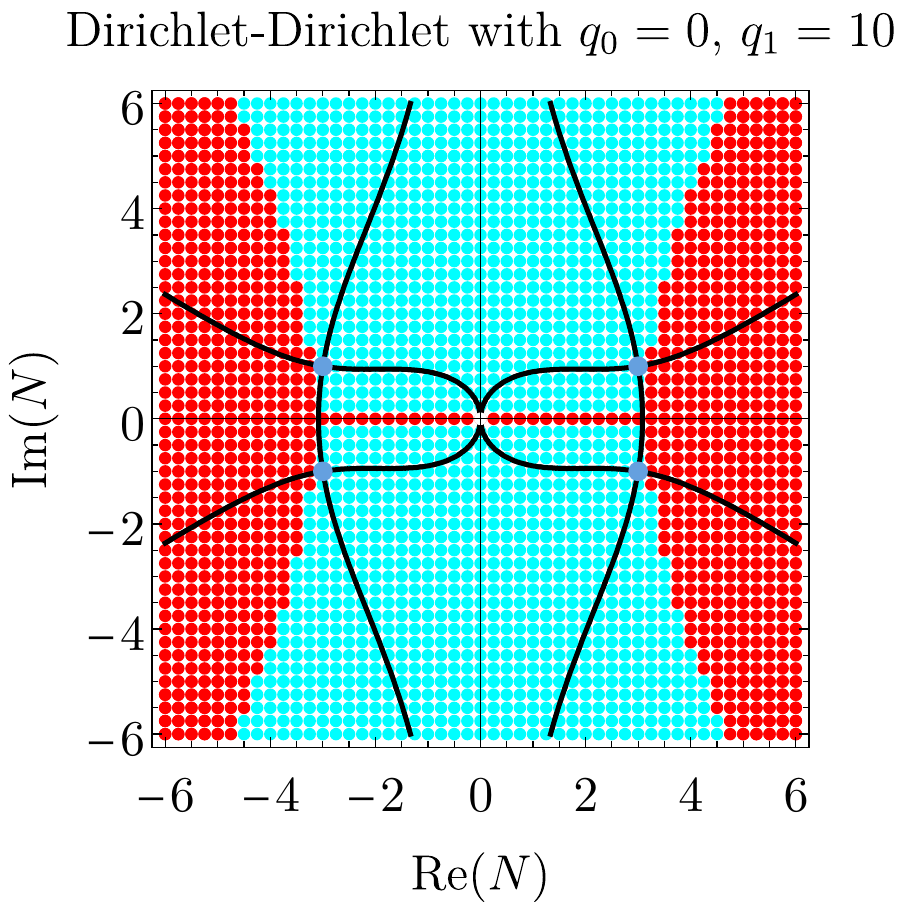} \hspace*{0.5cm}
		\includegraphics[width=0.45\textwidth]{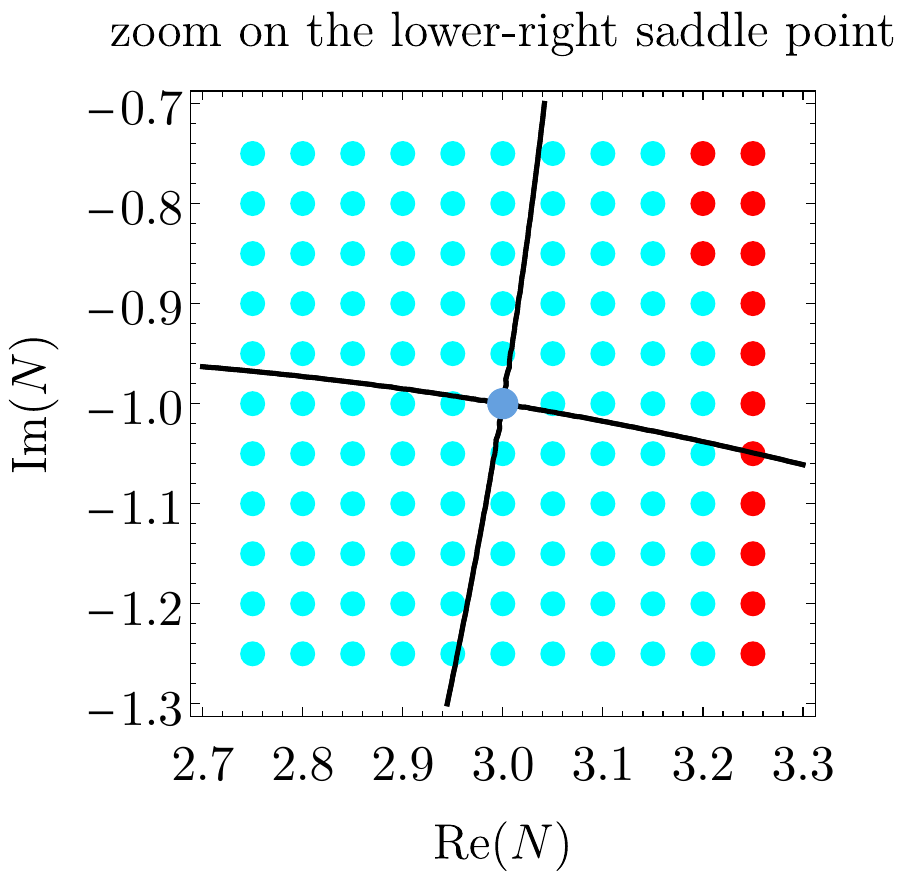}
		\caption{In the left panel, a numerical plot in the complexified lapse plane of the allowed (cyan) and disallowed (red) metrics satisfying initial and final Dirichlet conditions, as appropriate for a field space definition of the no-boundary wave function. Here $D=4$, $\Lambda=3$, $q(t=0)=0$, and $q(t=1)=10$. This time the saddle points (in blue) do not reside at the boundary of the allowable domain, but are surrounded by allowable metrics, as can be seen more clearly in the zoomed in version in the right panel.}
		\label{fig:DD4D}
	\end{figure}

	Summing over compact metrics means that we will impose an initial Dirichlet condition, namely $q(t=0)=0$, while retaining also a Dirichlet condition on the final hypersurface, $q(t=1)=q_1$. Our approach will follow the same steps as in the previous section. Once again, the path integral over the scale factor can be carried out by expanding around a background, $q(t)=\bar{q}(t)+Q(t)$, with $Q(0)=Q(1)=0$ but otherwise arbitrary. A background solution to the equation of motion \eqref{qeom} satisfying the Dirichlet boundary conditions is given by
	\begin{equation}
		\bar{q}_\text{DD}(t) = \frac{\Lambda}{3}N^2 t(t-1) + q_1t\,. \label{metricDD}
	\end{equation}
	The integral over $Q$ is a Gau\ss{}ian integral, which just changes the measure by a factor of $1/\sqrt{N}$, which we will ignore since we are ultimately interested in the semi-classical limit. This leaves an ordinary integral over the lapse,
	\begin{equation}
		\Psi_\text{DD}(q_1) = \int \dd N\,\exp[\frac{2\pi^2i}{\hbar}\left(\frac{\Lambda^2}{36}N^3+\left(3-\frac{\Lambda q_1}{2}\right)N-\frac{3q_1^2}{4N}\right)]\,.
	\end{equation}
	The saddle points of the integrand are given by
	\begin{equation}
		N_\text{SP} = \frac{3}{\Lambda}\left(\pm\sqrt{\frac{\Lambda q_1}{3}-1}\pm i\right)\,.
	\end{equation}
	There are $4$ saddle points, obtained by choosing the signs above independently. Their physical properties differ starkly \cite{Halliwell:1988ik,Feldbrugge:2017kzv,Feldbrugge:2017fcc}. The two saddle points in the lower half plane have an enhanced weighting $\exp(+\frac{1}{2\hbar}|{\mathcal I}(S^4)|)$ (where ${\mathcal I}(S^4)$ is the action on a $4$-sphere of radius $\sqrt{3/\Lambda}$), and perturbations around these geometries are stable. By contrast, the two saddle points in the upper half plane have a suppressed weighting $\exp(-\frac{1}{2\hbar}|{\mathcal I}(S^4)|)$ and unstable perturbations (as discussed in detail in \cite{Feldbrugge:2017mbc,DiazDorronsoro:2017hti}). The saddle points and the associated steepest ascent/descent contours are shown in the left panel of figure~\ref{fig:DD4D}. Based on this, we can explain the objection to defining the no-boundary wave function by a sum over compact metrics raised in \cite{Feldbrugge:2017mbc}: there exists a steepest descent contour from each of the saddle points in the lower half plane that links up with a saddle point in the upper half plane. Hence, any integration contour that includes a stable saddle point will necessarily also include an unstable saddle point, thus rendering the calculation inconsistent. Our aim here is to explore the structure of the space of metrics from the point of view of the LSKSW bound, to see if this has any impact on the definition of the no-boundary wave function.
	
	As before, we apply the numerical methods described in section \ref{subsec:methods}. Our results are shown in figure~\ref{fig:DD4D}. In these graphs, which show a grid of geometries in the complexified lapse plane, the blue dots represent the saddle points. Cyan dots indicate metrics that are allowable, while red dots show metrics that are not allowable according to the LSKSW criterion. There are at least two interesting features: the first is that the saddle points, in contrast to the definition in momentum space, do not reside on the boundary of the allowed domain, but rather are surrounded by well behaved geometries (see in particular the magnified plot in the right panel). This suggests that the momentum space and field space no-boundary wave functions cannot be exact Fourier transforms of each other, though in the semi-classical limit they may be closely related. The second feature is that the real-$N$ axis contains only disallowed geometries. These metrics are in fact exactly Lorentzian, as can be seen from \eqref{metricDD}. As such, they saturate the LSKSW bound, $\Sigma=\pi$. Thus, with our minisuperspace metric ansatz \eqref{metric}, this line divides the space of metrics into two separate halves, the upper- and the lower-half planes. Any contour of integration is cut off on this line, and hence the argument given above, namely that the saddles in the lower-half plane automatically link up with those in the upper-half plane, becomes moot. Hence, within the assumptions of the present model, one can define a no-boundary wave function as a sum over compact metrics after all, the contour of integration being restricted to the steepest descent contours in the lower half plane (i.e., the contours coming from $-i \infty$ and passing through either of the saddle points). If one desires a real wave function, as advocated by Hartle \& Hawking \cite{Hartle:1983ai}, then one can sum both contours with equal weight.
	
	A word of caution is required, however: our analysis above took place within a very restricted set of metrics (namely within minisuperspace). An important question is whether the no-boundary wave function could be defined as a sum over compact metrics in full generality. This question translates into the question of whether or not there exist allowable metrics with zero weighting that satisfy the Dirichlet boundary conditions $q(0)=0$, $q(1)=q_1$, for arbitrary $q_1>3/\Lambda$, and which can continuously link an integration contour stemming from a saddle in the lower-half plane to an integration contour stemming from a saddle in the upper-half plane. Answering this question is beyond the scope of the present work.
	
	We nevertheless make the conjecture that the configuration space of metrics will always be split into two disconnected regions of allowability containing the complex conjugated saddle points, as it appears to be the case for FLRW metrics studied here due to the uncrossable line of non-allowable real Lorentzian metrics. If this conjecture is proven true, then it would put on strong footing the assertion that the no-boundary wave function as a sum over compact metrics can indeed be well defined and stable.

	%%%%%%%%%%%%%%%%%%%%%%%%%%%%%%%%%%%%%%%%%%%%%%%%%%%%%%%%%%%%%%%%%%%%%%%%%%%%%%%%%%%%%%%%%%%%%%%%%%%%%%%
	
	\section{Discussion and conclusions}
	\label{sec:discussion}
	
	We have analysed minisuperspace quantum cosmological models in light of the Louko-Sorkin-Kontsevich-Segal-Witten criterion for allowing or discarding complex metrics. We find that in situations which are well understood, the criterion gives reasonable results. For instance, there is no obstruction for calculating transition amplitudes between classical configurations of the universe, though we found that some care must be taken in using the criterion on off-shell metrics with restricted functional freedom. 
	
	In more speculative settings, the power of the allowability bound becomes manifest: in particular, we find that simple examples of quantum bounces, i.e., quantum transitions between contracting and expanding universes, are not allowable. This raises the important question of whether quantum bounces are disallowed by quantum gravity in general. (Note that many studies of quantum bounces are based on working with the wave function alone and do not investigate the effective geometry that is implied.) It is certainly an interesting challenge to find an allowable complex metric mediating a quantum bounce. Conversely, one may ask if it is possible to prove that no quantum bounces can be allowable. Such a result would have far reaching consequences, as it would be a strong indication that the universe indeed had a beginning almost $14$ billion years ago.
	
	Regarding such a beginning, the no-boundary proposal still remains the best understood effective description to date. Here we find that when the no-boundary wave function is defined (in momentum space) as a sum over regular metrics, then its saddle points are allowable and moreover they sit on the edge of the domain of allowable metrics, in any spacetime dimension. Furthermore, we explicitly identified suitable integration contours in all dimensions. It is significant that this no-boundary path integral fits so well with the criterion on allowable metrics since it requires the use of complex metrics. We also found encouraging results regarding a definition of the no-boundary wave function in field space, as a sum over compact metrics. In minisuperspace, the Lorentzian metrics with real lapse and zero weighting divide the space of metrics into two halves. Restricting the sum to just one of those halves would in fact allow for a consistent definition, in which the stable saddle points of the integral are disconnected from unstable configurations. Whether such a division of the domain of metrics into two halves (one with positive and one with negative weightings) continues to hold for more general metrics is a crucial open question for future research.
	
	In this work, we showed at many instances that if the LSKSW criterion is applied, there are situations in which one cannot use (full) steepest descent contours as integration contours. Rather, at some point one is forced to deviate from the thimbles. If we stick to minisuperspace, then we have to cut the integration contours off; if we allow a larger class of metrics, we can presumably continue the contours within this larger class. It is not that we have no control over the thimbles, or that we do not trust the thimbles, it is just that they would have been the optimal integration contours, but one cannot use them in full for this purpose. Here, we should emphasise that we analyse the structure of the lapse integral assuming that the scale factor is already on-shell.\footnote{Let us point out here that many of our conclusions still hold even if $q$ is treated off-shell. In particular, all real Lorentzian homogeneous metrics are still on the boundary of the LSKSW bound, violating it. As such, configurations on the real-lapse line are still non-allowable and create a separation among allowable geometries. It is hard to comment on other off-shell configurations (both in $N$ and $q$) in more generality though, and this certainly deserves closer scrutiny in a future study. Evidently, the analysis of the (non-)allowability of saddle points, for instance such as those considered in the section on quantum bounces, remains unaffected by these considerations.} It then becomes clear that off-shell one may not need to cut off the integration contour; however, the contours are not thimbles anymore then, i.e., they cannot remain the steepest descent paths. This is a consequence of the LSKSW criterion; whether this consequence is desirable or not needs to be discussed. Yet, it also means that the cut-off points are not necessarily physical. These might merely imply that the integration contour has to leave the thimbles there.
	
	The success of the LSKSW criterion to date motivates asking whether and how it might fit into a more general quantum gravity framework. These aspects provide many avenues for future investigations. First, we note that the criterion may be seen as an instance of demanding well-defined semi-classical physics. This provides a direct link to our work on finite amplitudes \cite{Jonas:2021xkx} and also to the swampland program -- it was speculated both in \cite{Jonas:2021xkx} and in \cite{Hamada:2021yxy} that ultimately most swampland conditions might follow from a strict implementation of semi-classical consistency conditions. But are there more basic underlying principles, from which semi-classical consistency would follow as an approximation? And could the swampland distance conjecture \cite{Ooguri:2006in}, which limits the field range over which integrations can be performed, weaken the LSKSW criterion since it would modify the convergence condition for scalar fields? Second, in theories containing extra spatial dimensions, one may ask whether there are links between allowability in different dimensions. For instance, how do the allowability criteria in different dimensions mesh when higher-dimensional metric degrees of freedom actually count as fluxes or scalars from the lower-dimensional point of view? And third, is there a special meaning to the boundary of the allowed space of metrics in quantum gravity? Many physically useful metrics reside on this boundary, and a first step would be to try to determine the shape of this boundary. Are we compelled to live on this edge?

	%%%%%%%%%%%%%%%%%%%%%%%%%%%%%%%%%%%%%%%%%%%%%%%%%%%%%%%%%%%%%%%%%%%%%%%%%%%%%%%%%%%%%%%%%%%%%%%%%%%%%%%
	
	\vskip23pt
	\subsection*{Acknowledgments}
	
	We would like to thank Alex Maloney for a stimulating discussion.
	We gratefully acknowledge the support of the European Research Council (ERC) in the form of the ERC Consolidator Grant CoG 772295 ``Qosmology''.
	This publication was also supported by the Fields Institute for Research in Mathematical Sciences. Its contents are solely the responsibility of the authors and do not necessarily represent the official views of the Institute.
	We acknowledge the use of the Albert Einstein Institute's compute server ZERO for some of our numerical computations.
	This research made use of \texttt{Jupyter} \cite{soton403913}, \texttt{NumPy} \cite{Harris:2020xlr}, and \texttt{SciPy} \cite{Virtanen:2019joe}.
	\vskip23pt

	%%%%%%%%%%%%%%%%%%%%%%%%%%%%%%%%%%%%%%%%%%%%%%%%%%%%%%%%%%%%%%%%%%%%%%%%%%%%%%%%%%%%%%%%%%%%%%%%%%%%%%%
	%\newpage
	\phantomsection
	{\renewcommand{\baselinestretch}{0.9}
		\addcontentsline{toc}{section}{References}
		\bibliographystyle{JHEP2}
		\bibliography{references}}
	
\end{document}